\documentclass[aps,showpacs,prb,twocolumn,supperscriptaddress]{revtex4-1}
\usepackage{graphicx} 
\usepackage{dcolumn} 
\usepackage{bm} 
\usepackage{graphicx,color,epsfig,rotate}
\usepackage{amsmath,amssymb}
\usepackage{pifont} 
\pdfoutput=1

\begin{document}

\title{Dynamical scaling for underdamped strain order parameters \\ quenched below first-order phase transitions}
\author{N. Shankaraiah$^\ast$, Awadhesh K. Dubey$^\ast$, Sanjay Puri$^\ast$ and Subodh R.  Shenoy$^\dagger$}
\affiliation{$^\ast$School of Physical Sciences, Jawaharlal Nehru University, New Delhi 110067, India; \\ 
$^\dagger$TIFR Centre for Interdisciplinary Sciences, TIFR Hyderabad 500075,  India.}

\begin{abstract}
In the conceptual framework of phase ordering after temperature quenches below transition, we consider the underdamped Bales-Gooding-type 'momentum conserving' dynamics of a 2D martensitic structural transition from a square-to-rectangle unit cell. The one-component or $N_{\rm OP} =1$ order parameter is one of the physical strains,  and the Landau free energy has a triple well, describing a first-order transition. We numerically study the evolution of the strain-strain  correlation, and find that it exhibits dynamical scaling, with a coarsening length $L(t) \sim  t^{\alpha}$. We find at intermediate and long times that the coarsening exponent sequentially takes on respective values close to $\alpha=2/3$ and $\alpha=1/2$. For deep quenches, the coarsening can be arrested at long times, with $\alpha \simeq 0$. These exponents are also found in 3D.  To understand such behaviour, we  insert a dynamical-scaling ansatz into the correlation function dynamics to  give, at a dominant scaled separation, a nonlinear kinetics of the curvature $g (t) \equiv 1/ L(t)$. The curvature  solutions  have time windows of  power-law decays $g \sim 1/t^\alpha$, with exponent values $\alpha$ matching simulations, and manifestly independent of  spatial dimension. Applying this curvature-kinetics method to mass-conserving Cahn-Hilliard dynamics for a double-well Landau potential in a scalar $N_{\rm OP}=1$ order parameter yields exponents $\alpha = 1/4$ and $1/3$ for intermediate and long times. For vector order parameters with $N_{\rm OP} \geq 2$, the exponents are $\alpha = 1/4$ only, consistent with previous work. The curvature kinetics method could be useful in extracting coarsening exponents for other phase-ordering dynamics.
\end{abstract}

\maketitle

\vskip 0.5truecm

\section{Introduction}

Interacting systems such as magnets, binary fluids, liquid crystals, and quantum spin models can be probed by the dynamical evolutions of order parameters, after temperature or coupling-constant quenches below transition \cite{R1, R2, R3, R4, R5, R6,R7,R8,R9,R10,R11,R12,R13}. Such phase ordering can be quantitatively described by a time-dependent, two-point, order-parameter correlation $C(\vec R,t)$, that can exhibit {\it dynamical scaling}, dependent on  space and time through a single scaled variable\cite{R1,R5,R6,R7,R8,R9,R10,R11,R12,R13} ${\bar R} \equiv |\vec R| /L(t)$, with consequent data collapse onto a single scaled curve $G(\bar R)$. The coarsening length $L(t)$ is a measure of the typical spacing between domain walls separating competing order parameter (OP) phases. It can increase as $L(t)\sim t^{\alpha}$, where the exponent $\alpha$ is independent of  material parameter values, but could depend on  the nature of the OP dynamics, the number of components of the order parameter $N_{\rm OP}$, the number of competing low-temperature variants $N_V$, and the spatial dimension $d$. There can be a sequential appearance of different exponents, during coarsening \cite{R1}.

In various dynamic models, the exponents $\alpha$ have been estimated by heuristic arguments, energy dissipation matchings, Gaussian fluctuations of domain wall profiles, self-consistent correlation function dynamics, and through numerical simulations \cite{R1,R5,R6,R7,R8,R9,R10,R11,R12}. The Allen-Cahn relaxation equation \cite{R1} for  a nonconserved OP and the Cahn-Hilliard equation \cite{R1,R4} for a  locally conserved OP are familiar models, both with a single time derivative, and typically use a double-well Landau free energy describing a second-order transition. It is of much interest to explore other phase-ordering dynamics;  and to develop systematic methods of estimating coarsening-length exponents. 

Solid-solid  structural transitions have  $N_{\rm OP}$ strain-tensor components as the order parameters \cite{R14,R15,R16,R17,R18,R19,R20,R21}, with Landau free energies having $N_V$  competing minima, for the $N_V$ different unit cells, or 'variants'. The high-temperature, high-symmetry crystal structure is ``austenite,'' and the low-temperature, low-symmetry structures are ``martensite.'' Martensitic transitions can be described by a Bales-Gooding or BG-type strain dynamics \cite{R19} that has several features \cite{R19,R20}, which differ from the more familiar magnetism-inspired phase orderings. Firstly, the Landau free energy \cite{R16} can have {\it triple} wells in the OP, describing a first-order phase transition, with a minimum also at zero values of the OP. Secondly, the dynamics is  {\it underdamped}, with a Newtonian inertial term or  double time  derivative, describing acceleration of the order parameter. Thirdly,  there is global momentum conservation, with the single time derivative damping  term, suppressed at  long wavelengths\cite{R19,R20}. Fourthly, with a 2D dynamics \cite{R20} generalized  from \cite{R19} 1D, the order parameters  have an additional power-law anisotropic interaction, coming from an elastic  St-Venant compatibility constraint \cite{R15,R16}, as used in several contexts \cite{R21}. 

Monte Carlo simulations in 2D of a discretized strain-pseudospin martensitic model Hamiltonian with power-law anisotropic (PLA) interactions show interesting evolutions under a temperature quench. For example, for successive  quenches approaching the transition from below, the conversion time from seeded austenite evolving  to martensite domains rises sharply, with these time delays caused by entropy barriers \cite{R18}. Clearly, martensites with continuous-strain dynamics \cite{R19,R20} are worth examining, in the framework of phase ordering ideas \cite{R1}. We here focus on effects of the triple well Landau term in the underdamped dynamics, and suppress the power-law anisotropic interaction, which will be considered in a subsequent publication. 

In the first part of this paper, we apply phase-ordering ideas to the BG strain dynamics, with  $d=2$, $N_{\rm OP}=1$, and $N_V=2$, in a sixth-order Landau polynomial in the scalar OP strain. We consider only Landau and Ginzburg terms in the OP dynamics to numerically determine the dynamic structure factor or OP-OP correlation, finding dynamical scaling in a coarsening length $L(t)\sim t^{\alpha}$. The exponent takes on sequential values such as $\alpha = 2/3, 1/2$ over time windows, whose widths depend on the quench temperature $T$.  For deep quenches, there is an exponent $\alpha =1/3$, and a final $\alpha \simeq 0$  flattening to a constant, analogous to 'coarsening arrest' \cite{R22}. These 2D results are found to persist, for 3D.

In the second part of the paper, we use a scaled form of the underdamped OP dynamics, to obtain a dynamics for the OP-OP correlation $C (R, t)$. Inserting the dynamic scaling form, $C= G ( R/ L(t))$, yields a nonlinear, underdamped  kinetics for the curvature or inverse coarsening length $g (t)  \equiv 1/ L(t)$, with coefficients evaluated at dominant coarsening-front separation. Here, to close what would otherwise be an infinite hierarchy, a  spatial average of internal domain-wall factors is made, reducing the correlation between the chemical potential and order parameter, to the OP-OP correlation $G (gR)$.   The curvature kinetics solutions from balancing kinematic and force terms are simple power-law decays  $g (t) \sim 1/ t^\alpha$ in sequential time windows, showing exponents $\alpha =2/3,1/2$ values, independent of $d$, in agreement with simulations. For deep quenches, a toy model including higher powers of the curvature, explains the $\alpha=0$ coarsening arrest, as a metastable trapping of curvature to a nonzero value.

As a check, the  curvature kinetics method is applied to Cahn-Hilliard dynamics, yielding  the well-known \cite{R1,R8,R9}  values of $\alpha=1/3$  for a scalar order parameter $N_{\rm OP} =1$, and $\alpha = 1/4$ for  vector order parameters with $N_{\rm OP} \geq 2$, all independent of $d$.

The plan of the paper is as follows. In Sec. II, we state the Cahn-Hilliard and Bales-Gooding types of OP dynamics, and scale to absorb  all, or most, of the OP  temperature dependencies. Section III  defines the OP-OP correlation functions and their dynamical scaling. Section IV shows the coarsening textures, numerically demonstrates dynamical scaling of the two-point correlation function, and states the obtained exponents.  The second part of the paper, starting  in Sec. V, obtains the correlation function dynamics, and inserts a dynamical-scaling ansatz. Section VI extracts the curvature kinetics, and predicts power-law exponents $\alpha$ that match the numerics. Finally, Sec. VII contains a discussion and comments on future work.  Details are given,  in Appendix A, of the closure approximation; in Appendix B, of coefficient signs in the curvature kinetics; in Appendix C, of estimation of coarsening exponents.

\section{Different order parameter dynamics}

\subsection{ Cahn-Hilliard type dynamics}

The simplest order parameter dynamics is the purely relaxational or over-damped Allen-Cahn equation \cite{ R1}  for a nonconserved order parameter that says the damping force balances the chemical-potential driving force from the free energy, $\dot e \sim -\partial F /\partial e$.  Another dynamics is the Cahn-Hilliard equation \cite{R4}  describing the evolution  of a conserved order parameter $e (\vec r, t)$ that could be  a mass-concentration density or a magnetization, 
$$ \gamma  {\dot e}(\vec r, t)  =  (-\vec \nabla^2) \left[ - \frac{\partial F ( e)}{\partial e (\vec r, t)} \right].  ~~(2.1)$$
This can be written as a continuity equation
$$ \partial e (\vec r, t)/\partial t + \vec \nabla \cdot {\vec j}(\vec r, t) ,~~~(2.2a)$$
where the diffusion current density is driven by spatial gradients of the chemical potential  
$${\vec j} = - D \vec \nabla \mu (\vec r, t), ~~\mu \equiv  \frac{\partial F ( e)}{\partial e (\vec r, t)}.~~ (2.2b)$$
The diffusion constant is the inverse friction coefficient, $D= \gamma^{-1}$.
Since ${\dot e} (\vec k, t) \sim - k^2 \mu (\vec k,t)$,  the  uniform order parameter $k \rightarrow 0$ is independent of time, i.e., the spatial average of the order parameter is conserved.

The free energy in terms of the OP is typically taken as a Ginzburg or gradient term, plus a Landau term for a second-order transition, $F = F_G + F_L$, 
where $F_G = E_0  \sum_{\vec r} \xi_0 ^2 (\vec \nabla e)^2$, $F_L = E_0 \sum_{\vec r} [ \epsilon e^2 + e^4 /2]$. Here,  $\xi_0$ is a bending length, $\epsilon \equiv (T -T_c)/T_c$, and $E_0$ is an energy density. 

Re-defining space and time variables to make them dimensionless, in units of  a numerical grid length $a_0$, and a  chosen time unit, 
$r \rightarrow  a_0  r; ~~~  t  \rightarrow(\gamma /2 E_0) t,$											
  the Cahn-Hilliard equation of (2.2) becomes 
$$
\frac{\partial e(\vec r,t)}{\partial t} = {\vec \nabla}_{\vec r}^2 ~  \mu(\vec r,t),~~~(2.3a)
$$
where the chemical potential has Landau and Ginzburg terms,

$$\mu =\mu_L - \xi_0^2  {\vec \nabla^2} e~~(2.3b),$$
where $\mu_L \equiv (1/2) \partial f_L /\partial e =  -(|\epsilon|  -e^2) e$.

\subsection{Bales-Gooding type dynamics}

 \begin{figure*}[ht]
\begin{center}
\includegraphics[height=4.5cm, width=15cm]{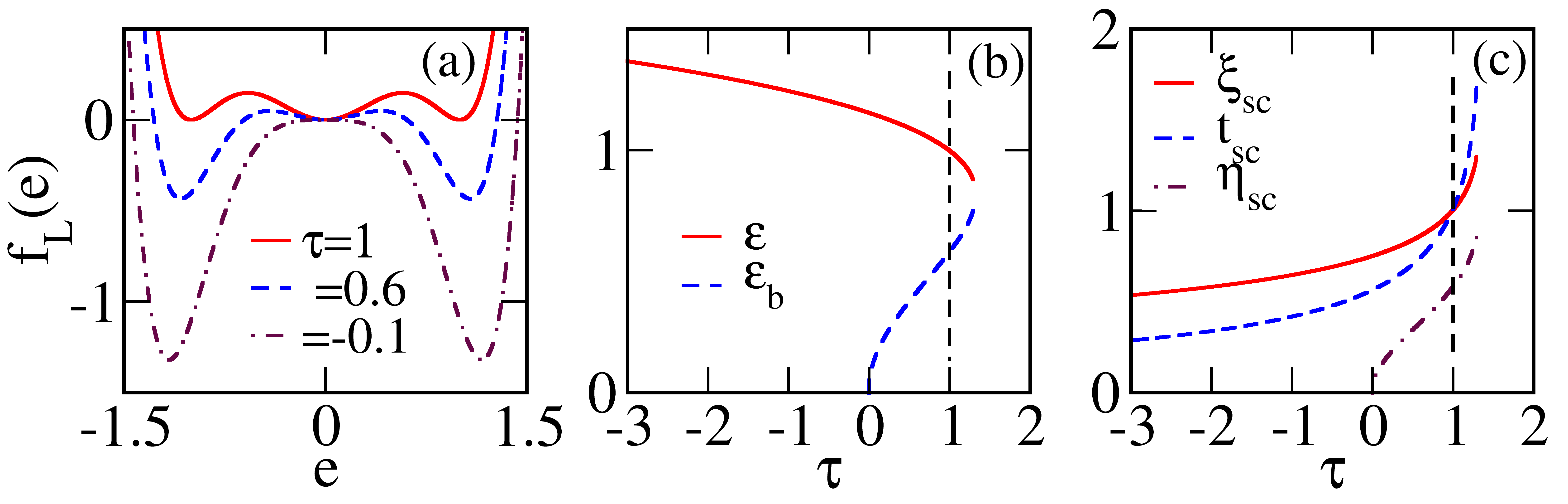}
\caption{(Color online) {\it BG case temperature dependent parameters:}  (a) Landau free energy density $f_L$  vs order parameter $e$, for various temperatures $\tau$. There are  well-minima at $\bar \varepsilon$, and  barrier-maxima at ${\bar \varepsilon}_b$. (b) Mean-field order parameter $\bar \varepsilon$ and barrier ${\bar \varepsilon}_b$, vs $\tau$. (c) Scaling length $\xi_{\rm sc}$, scaling  time $ t_{\rm sc}$, and residual $T$-dependence of $\eta_{\rm sc} (\tau)$, vs $\tau$.}
\label{Fig.1}
\end{center}
\end{figure*}

Bales and Gooding  (BG) have used a Lagrangian formalism to obtain \cite{R19} a continuous-strain, underdamped, momentum-conserving dynamics in 1D, for a one-component order parameter, or  $N_{\rm OP}=1$. 
For 1D, there is only one type of strain  $e= \partial u (x)/\partial x$ or gradient of displacement $u(x)$, that is, the OP. The free energy  $F=E_{0} \sum_{x}f$ where the free energy density $f =f_L+f_G$ is a sum of a {\it triple}-well Landau term $f_L = e^6- 2 e^4+\tau e^2$, and a Ginzburg term  $f_G \sim \xi_0 ^2 (\partial e/\partial x)^2$.

The Lagrangian density is $\rho_0 \dot e^2-f$ where in the ``kinetic-energy'' term,  $\rho_0$ is the mass of the unit cell of volume ${a_{0}}^{d}$, and the ``potential-energy'' is  $f$. With a Lagrangian minimization, and adding a Rayleigh dissipation term $\sim -\gamma \dot e$, one  gets \cite{R19}
$$
 \rho_0 \ddot e (x,t) = \left(\frac{-\partial^2}{\partial x^2}\right) \left[\frac{-\partial F}{\partial e (x,t) }-\gamma \dot e (x,t)\right].~~(2.4)
 $$
 
 Note that the long-wavelength $k \rightarrow 0$ limit enforces global conservation of the total system momentum, with nonzero damping only for $\vec k \neq 0$ internal momenta.

For higher spatial dimensions $d=2$ and $3$, there are {\it multiple} strain components,  describing the physical shear, compression and ``deviatoric'' or rectangular, distortions of the unit cell. A subset of the physical strains are the $N_{\rm OP}$  order parameter components that enter the nonlinear Landau free energy. The remaining  non-OP physical strains are linked to the OP strains by \cite{R15,R16} ``compatibility constraints'' that ensure  the distorted unit cells fit together in a smoothly compatible way, without dislocations. A {\it  constrained} minimization yields an OP-OP effective interaction with an elastic constant prefactor $A_1$,
that is, power-law and anisotropic, inducing preferred diagonal domain-wall orientations \cite{R15,R16}. We will throughout, set $A_1=0$, and consider these compatibility-induced interactions, elsewhere. 

For a square-to-rectangle transition,  the OP is the deviatoric strain written as $e$, and as before,  the free energy 
$F = E_0 \sum [f_L + f_G ]$.  
The   triple-well Landau term as shown in Fig 1(a), has  minima at $e=0$, $e=\pm {\bar \varepsilon}(\tau)$,
$$
 f_L = [(\tau-1)e^2+e^2(e^2-1)^2 ] ,~~(2.5)
$$
with a scaled temperature defined as
$$
 \tau(T) \equiv \frac{T-T_c}{T_0-T_c}. ~~ (2.6a)
$$
The minima are at $e = \pm {\bar \varepsilon} (\tau)$, where 
$${\bar \varepsilon} ^2 \equiv \frac{2}{3} [1+\sqrt{1-3\tau/4}],~~(2.6b)$$
while the barriers between the austenite and martensite wells are at  ${\bar \varepsilon}_b (\tau)^2 \equiv \frac{2}{3} [1-\sqrt{1-3\tau/4}]$.
Here, just below $T=T_0$, or $\tau(T_0)=1$,  when the triple wells are degenerate, the  OP jumps from zero to unity $\bar \varepsilon (1)  = 1$. Below  the austenite spinodal $T=T_c$ or $\tau(T_c)=0$, the barriers  vanish, $ {\bar \varepsilon}_b (\tau=0) =0$, and the metastable austenite minimum at $e=0$ disappears,  (when the martensite minima are at  $\bar \varepsilon (0)  = \pm \sqrt{4/3}$). At zero temperature, $\tau (0) = -T_c/ (T_0 - T_c)$ (see Fig 1).

The Ginzburg term is 

$$F_G = E_0 \sum \xi^2 _0 (\vec \nabla e)^2 , ~~(2.7)  $$
where $\xi_0$ is an OP bending length scale.

The underdamped dynamics is \cite{R20},
$$
 \rho_0 \ddot e (\vec r,t) = c_0^2 \vec \nabla ^2 \left[ \frac {\partial F}{\partial e}+ \gamma \dot e \right], ~~ (2.8)
$$
where $c_0=\frac{1}{2}$ is a  normalization. Here with a compatibility term  $f_C \sim A_1 =0$ suppressed,  $f = f_L + f_G$, and 
$$
 \frac{\partial f}{\partial e({\vec r},t)}=\frac{\partial f_L}{\partial e} -2  \xi_0^2 {\vec \nabla}^2 e . ~~(2.9)
$$

Re-defining  space and time variables  as before,
$r \rightarrow  a_0 r; ~~~  t  \rightarrow (\gamma /2E_0)t$,
 (2.8) becomes
$$
 \Lambda \frac{\partial^2 e(\vec r,t)}{\partial t^2}= {{\vec \nabla}_{\vec r}}^2 \left[\mu(\vec r,t)+\frac{\partial e(\vec r,t)}{\partial t} \right],~~(2.10)
$$
where the dimensionless $\Lambda \sim \gamma^{-2}$ is an inverse-damping squared and, with  the unit-cell length $a_0 =1$, is

$$
\Lambda \equiv \left[ \frac{2 E_0  \rho_0 }{c_0^2} \right] \frac{1}{\gamma^2}.~~~(2.11)
$$
 The  chemical potential is
$$
 \mu(\vec r,t) = \mu_L - {\xi_0}^2 {\vec \nabla}_{\vec r}^2  e(\vec r,t),~~ (2.12)
$$
where  $\mu_L  \equiv (1/2) \partial f_L / \partial e = e[\tau - 4 e^2 +3 e^4]$. 

\subsection{Scaling out the OP $T$-dependence}

As is well-known \cite{R1,R2,R4}, the  CH dynamics can be cast in $T$-independent form, by scaling the OP by its Landau-minimum value, and
introducing $T$-dependent length and time scales, as
$$ e \rightarrow {\bar \varepsilon}~ e~; r \rightarrow r~ \xi_{\rm sc} (T);~ t \rightarrow t ~t_{\rm sc} (T),~~(2.13)$$
where ${\bar \varepsilon} = |\epsilon|^{1/2}$.  The kinetic term on the left side  of (2.3)  then has a factor $\xi_{\rm sc}^4 / t_{\rm sc}$, while the Landau term on right side has a factor $\xi_{\rm sc}^2 {\bar \varepsilon}^2$. Setting both equal to unity, the scaling length is seen to be the Ginzburg-Landau correlation length $\xi_{\rm sc} = 1/{\bar \varepsilon}= 1/|\epsilon|^{1/2} = \xi_{GL}(T)$, while
the scaling time is $t_{\rm sc} = 1/{\bar \varepsilon}^4 = 1/|\epsilon|^2$. The OP-scaled CH dynamics is then in the $T$-independent form, 
$$\partial e/ \partial t = {\vec \nabla^2} \mu_L - \xi_0 ^2  {\vec \nabla^4} e, ~~(2.14)$$
where $\mu_L  = -e f_0(e)$, with a scaled factor $ f_0 = (1-e^2)$, that vanishes in the bulk.

For the BG case, the OP-scaling of (2.13)  yields factors of the same type $(\xi_{\rm sc}^2/t_{\rm sc})^2$, and $\xi_{\rm sc}^2/ t_{\rm sc}$, for the inertial and damping terms, respectively, while the Landau term has a factor $\xi_{\rm sc} ^2  {\bar \varepsilon}^4$. Setting these to unity, 
$\xi_{\rm sc} = 1/ {\bar \varepsilon}^2 (\tau),~ t_{\rm sc} = 1/ {\bar \varepsilon}^4 (\tau)$.  The OP-scaled BG dynamics  without compatibility interactions is
$$\Lambda \partial^2 e/ \partial t ^2 = {\vec \nabla^2} [ \mu_L + \partial e /\partial t ] - \xi_0 ^2 {\vec \nabla^4} e , ~~(2.15)$$
where $\mu_L = -e f_0(e)$ with $f_0 = 3  (1- e^2)(e^2 - \eta_{\rm sc} (T) )$. Thus for a first-order transition, scaling the OP by its Landau value still leaves behind a  residual temperature-dependence,  through

$$\eta_{\rm sc} (T) \equiv \tau /3 {\bar \varepsilon}^4, ~~ (2.16)$$
that is negative for $\tau <0$ below the spinodal, and for $\tau >0$ is  essentially the (positive)  ratio of barrier height to well depth,  $\eta_{\rm sc} = [ {\bar \varepsilon}_b (\tau) / ~{\bar \varepsilon}(\tau)]^2$ [see Fig 1(c)].

For the numerical simulations of Sec. IV, we will use the {\it unscaled } or $T$-dependent forms (2.3), (2.10), and only later multiply the curvature-evolution data by the scaling lengths and times.
For the theoretical analysis of Sec. V, we will use the ``OP-scaled'' forms (2.14), and (2.15).

\section{Strain correlations and dynamical scaling}

In this section, we define the OP-OP and related correlations, and their dynamical scaling forms.

For a one-component  martensitic-strain order parameter $e(\vec r,t)$, we consider a two-point correlation between OP's at $\vec r=\vec R+\vec r_0$, and $\vec {r^{\prime}}=\vec r_0$ on a $d$-dimensional lattice, and at equal times $t$. With an average over all origins ${\vec r_0}$ and over many runs,  OP-OP correlations are dependent only on the separation
$\vec R = \vec r - \vec {r ^{\prime}}$,
$$
 C(\vec R,t)=  \langle e(\vec r,t) e(\vec {r^{\prime}}, t) \rangle, ~~(3.1a)
$$
$$
=\frac{1}{N} \sum_{r_0} <e({\vec R} +{\vec r}_0,t) e({\vec r}_0,t)> .~~(3.1b)
$$

With a Fourier expansion 
$e(\vec r,t)=\frac{1}{\sqrt N} \sum_{\vec k} e^{i \vec k \cdot \vec r} e(\vec k,t)$,
we get
$$
 C(\vec R,t)= \sum_{\vec k}  S (\vec k,t) e^{i \vec k \cdot \vec R}, ~~(3.2a)
$$
 where the time-dependent structure factor is
$$
 S (\vec k,t)= \langle |e(k,t)|^2 \rangle. ~~(3.2b)
$$
Since the OP is real, its Fourier coefficients $e(\vec k,t)^{*}= e(-\vec k,t)$ and so $S(\vec k,t) =S(-\vec k,t)$.

Another correlation that enters is the chemical potential-order parameter or  $\mu$-OP  correlation:
$$
 C^{(\mu)}(\vec R,t)=  \langle \mu(\vec r, t) e(\vec {r^{\prime}}, t) \rangle~~ (3.3a)
$$
$$
 = \left \langle \frac{1}{N}\sum_{r_0} \mu({\vec R}+{\vec r}_0,t)e({\vec r}_0,t) \right \rangle. ~~~(3.3b)
$$
With a Fourier expansion $\mu(\vec r,t)=\frac{1}{\sqrt N}\sum_{\vec k} e^{i \vec k \cdot \vec r} \mu(\vec k,t)$, we get
$$
 C^{(\mu)}(\vec R,t)=\sum_{\vec k}  S^{(\mu)}(\vec k,t) e^{i \vec k \cdot \vec R}, ~~~~(3.4a)
$$
where only the symmetric part  survives, and  the $\mu$-OP  Fourier correlation is the real part,
$$
  S^{(\mu)}(\vec k,t)=Re \langle \mu(\vec k,t) e(\vec k,t)^{\ast} \rangle. ~~(3.4b)
$$
 Since  the anisotropic compatibility interaction is switched off, the system is isotropic, and we can  work throughout with averages $<...>$ that now include angular averages, so correlations become
 $C(\vec R,t) \rightarrow C(R,t)$  and $S(\vec k,t) \rightarrow S(k,t)$,$S^{(\mu)} (\vec k,t) \rightarrow S^{(\mu)} (k,t)$ where $R \equiv |\vec R|$,  and $k \equiv |\vec k|$.
 
The OP-OP correlation $C(R,t)$ for a given time  $t$ will show a fall-off in separation $R$; while at a given separation $R$, it will increase with time.                                                        Since $\sim L(t)$ is the scale of the OP-correlation region, it will also increase with $t$.  As  $L(t)$ is also the separation of the OP-dips at domain walls, these strain  patterns must   {\it coarsen} with time. 

Dynamical scaling says that \cite{R1,R13} (i) the time $t$ enters only through the length $L(t)$;  (ii) the separation  $R$ appears only in scaled form as $\bar R \equiv  R/ L(t)$, [and in Fourier space, the wave vector $k$ appears only as $\bar k \equiv kL(t)$].  
Further, in a common assumption, (iii) the coarsening length scale grows as a power law $L \sim t^\alpha$, which is plausible \cite{R1}, but here is  justified.
The OP-OP correlation is then
$$
 C( R,t)/C(0,t) = G( R/L(t)) \equiv G(\bar R). ~~~(3.5)
$$

 We find later that the zero-separation values,  in  a few hundred time steps, on the onset of dynamical scaling, are insensitive to time, $C(0,t) \simeq C(0, t_{\rm onset})$, so  we henceforth suppress the constant denominator. 

In Fourier space,  the scaling behaviour is
$$
 S(k,t)  =  L^d \chi ( k L(t)) \equiv L^d \chi (\bar k).~~~(3.6)
$$

 For sharp domain walls, of widths small compared to $L(t)$, Porod's law holds for the Fourier space structure factor \cite{R1,R7}
$$
  \chi (\bar k) \simeq 1/[\bar k ]^{d+N_{\rm OP}}.~~~(3.7)
$$
For  $N_{\rm OP}=1$, we have  $\chi \simeq 1/(kL)^{d+1}$, and  a log-log plot of $S(k,t)$ versus $k$ will be a straight line with slope 
$-(d+1)$,  with the length $L(t)$ extracted from the intercepts $\ln \chi=-(d+1)\ln k-\ln L(t)$. Then with this scaling length, replots of $\ln \chi$ versus $\ln(\bar k)$, as well as $G(\bar R)$ versus $\bar R$, will show {\it data collapse} of different-time curves.

The coarsening curvature is defined \cite{R1} as $(d-1)/L(t)$, but we will simply work with an {\it inverse} length, or
$$
g(t) \equiv 1/ L(t), ~~~(3.8)
$$
and call this the ``curvature,'' as it indeed is, for $d=2$. Thus the scaled variables are
$$
\bar R \equiv g(t) R~; \bar k \equiv k / g(t).~~~(3.9)
$$
We later show that the correlation dynamics results in a nonlinear kinetics for the curvature $g (t)$,
that has power-law solutions $g \sim 1/ t^\alpha$, explaining the observed coarsening-length behaviour $L\sim t^\alpha$.

\section{Numerical results on dynamical scaling }

In this section, we present numerical work, showing domain-wall (DW) textures, demonstrating dynamical scaling, and extracting exponent behaviour of coarsening curvatures. 
In an actual  experimental quench, the physical temperature is changed suddenly, and the coarsening probe  is followed in physical time. Any universal curve independent of  quench temperatures $T$, is obtained only by later scaling  the measured physical values, by  factors dependent on $T$. As mentioned,  we mimic this  experimental procedure in simulations by using the {\it unscaled} forms of both the CH and BG dynamics in (2.3)and (2.10); only then doing scaling on the numerical data obtained. 

\begin{figure}[ht] 
\begin{center}
\includegraphics[height=3.5cm, width=8cm]{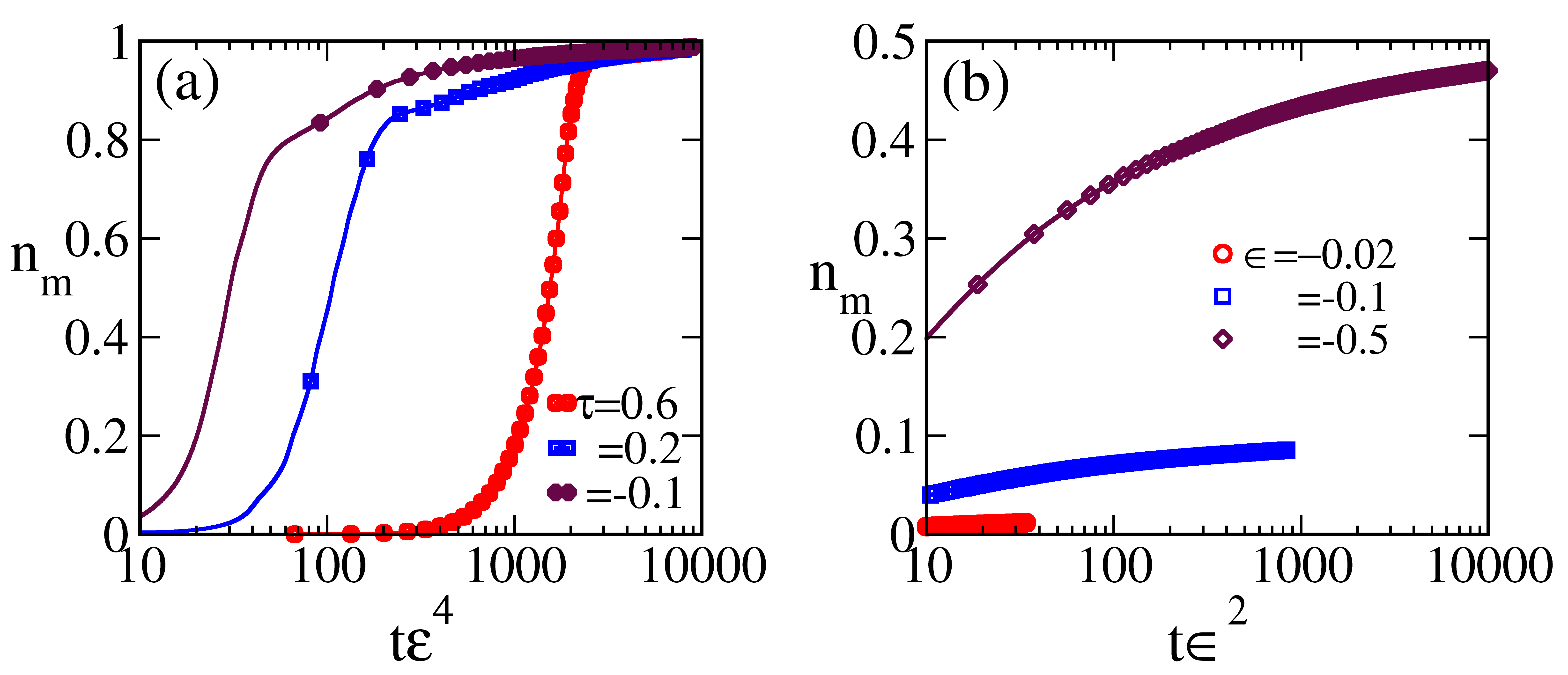}
\caption{(Color online) {\it Ordered-fraction evolutions:} (a) BG case martensitic-ordered fraction $n_m$ versus  scaled time $t {\bar \varepsilon }(\tau) ^4$, for  quenches $\tau= -0.1, 0.2, 0.6$. (b)  CH case ordered fraction $n_m$ versus  scaled time $t {\bar \varepsilon}^4 = t ~\epsilon (T) ^2$, for  quenches to $\epsilon\equiv (T/T_c -1) =  -0.5, -0.1, -0.02$. Critical slowing down is seen close to $T_c$, with ordered-fractions rising only slowly, within the holding-time $t < t_h$. }
\label{Fig.2}
\end{center}
\end{figure}

The initial  high-temperature state for both the CH and BG cases, corresponds to a {\it  single} well, with a minimum at $e(\vec r, t=0) =0$, plus a few local fluctuations of the ordered  variants. A numerical ``quench'' then corresponds to evolving at some  suddenly lower temperature $T$. There will be an early-time regime where the ordered phases expand, to crowd out the initially dominant $e=0$ background as a DW vapour phase;  while at intermediate and long times, there will be only competing variants, separated by domain walls, as a DW liquid phase.

\begin{figure}[ht]
\begin{center}
\includegraphics[height=3cm, width=8cm]{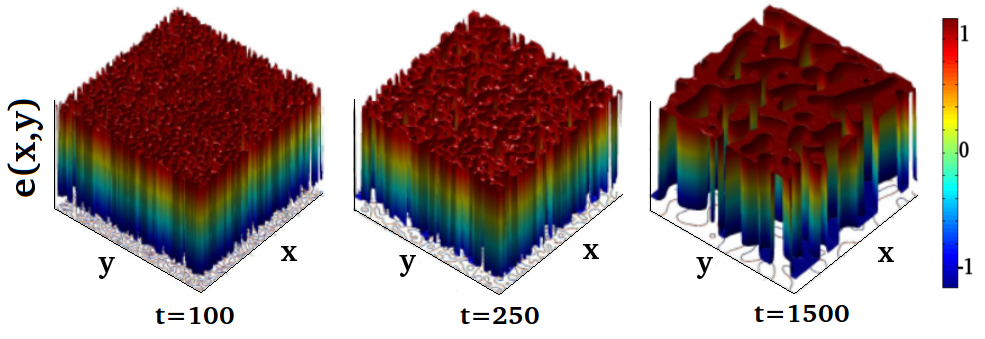}
\caption{(Color online) { \it Relief plot of  coarsening:} Snapshots for various times, showing  coarsening of the OP strain $e ({\vec r}, t)$ after a quench, for system size $512^2$.Positive/negative/zero strains are red/blue/green. See cross-sectional slices.The top of the relief plot shows positive (red) regions, plunging down through valleys, to reach complementary negative (blue) values, passing through domain walls regions,  where order parameters go through zero (green).}
\label{Fig.3}
\end{center}
\end{figure}

We  thus start with a dilute set of  initial seeds of both martensite variants equally,  in a sea of $e=0$ austenite. The martensite conversion fraction $ n_m (t) \equiv \frac{1}{N} \sum_{\vec r} e^2(\vec r,t) / | \bar \varepsilon| $ is initially nonzero only at  $2\%$  random sites surrounded by  $2 \times 2$ unit cells where $e=\pm {\bar \varepsilon}$, with $e=0$ elsewhere, and so $n_m (0) = 0.08$. 

\begin{figure*}[ht]
\begin{center}
\includegraphics[height=13cm,width=17cm]{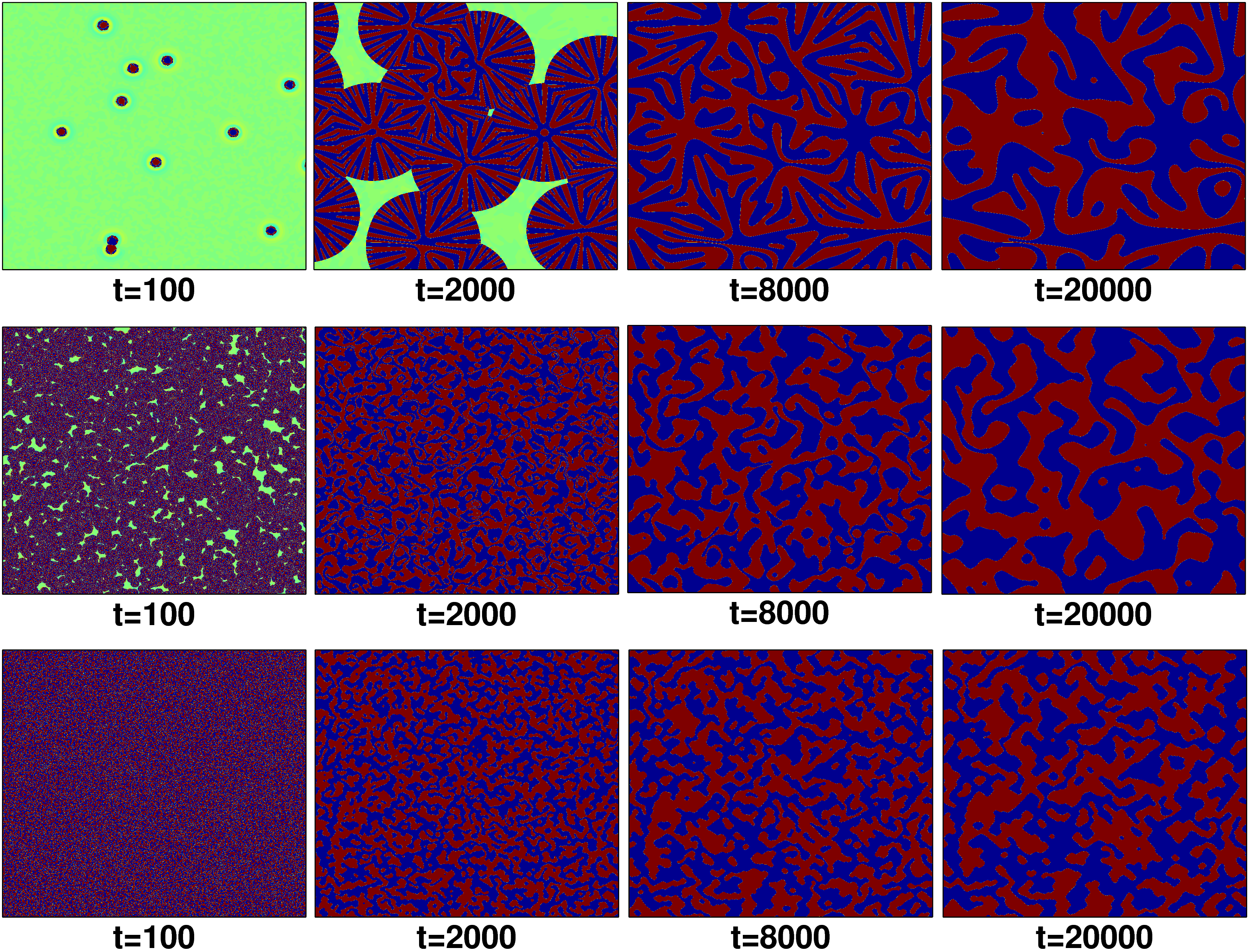}
\caption{(Color online) {\it Contour-plots of strain  evolutions:} Snapshots of $e (\vec r, t)$  textures, held at fixed quench temperatures, up to  a holding time $t_h = 20,000$  for system size $8192^2$; the pictures are zoomed in to an area of $4096^2$. (a) First row:  shallow quench to $\tau=0.6$, or $\Delta \tau = -0.1$; (b) Second row: moderate quench to $\tau=0.2$, or $\Delta \tau = -0.5$; (c) Third row: deep quench to $\tau=-0.1$, or $\Delta \tau =-0.8$, when coarsening seems to slow, suggesting a glassy state.}
\label{Fig.4}
\end{center}
\end{figure*}

We use an Euler-discretized dynamics of  (2.3), (2.10) to find the OP evolution of $e(\vec r,t)$, focusing on $\Lambda =1$. We take time steps $\Delta t = 0.05$;  and spatial  derivatives $\nabla_\nu$ as finite-difference operators $ (a_0)^{-1} \Delta_\nu$ on a square  unit lattice. For  wave vectors $\vec k$ in the Brillouin zone, $\Delta_\nu  \rightarrow 2 \sin (k_\nu a_0 /2)$ , with the grid scale set to $a_0 =1$. A fast Fourier transform yields  the Fourier coefficient $e (\vec k, t)$; and hence the angularly averaged dynamic structure factor  $S( k,t)=<|e(\vec k, t)|^2 >$ of (3.2). (To focus on DW time scales, a ``strain-hardening'' procedure is used \cite{R9}.) Run averages are taken, over $N_{run} =5$. A reverse FFT yields the  OP-OP correlations $C(R, t)$  of (3.1).  The quench  temperature $T$  is held fixed, for a holding time $t_h = 20,000$ steps. The 2D system is of size $ L_{\rm sys} ^2 =8192^2$. 

Figure 2(a) shows  the  BG case single-run martensite or ordered-fraction $n_m(t)$ versus scaled time $t \bar \varepsilon^4(\tau)$.  For  temperatures just below  $T =T_4$ or $\tau = \tau (T_4) =0.74$ the martensite fraction rises  slowly towards unity, with larger delays, closer to $T_4$. Above $T_4$ there is no conversion at all, as the initial martensite seeds dissolve  back into the four-fold symmetry austenite. We will quench to temperatures sufficiently below $T_4$ so conversion delays are small. 

\begin{figure}[ht]
\begin{center}
\includegraphics[height=6.5cm, width=8cm]{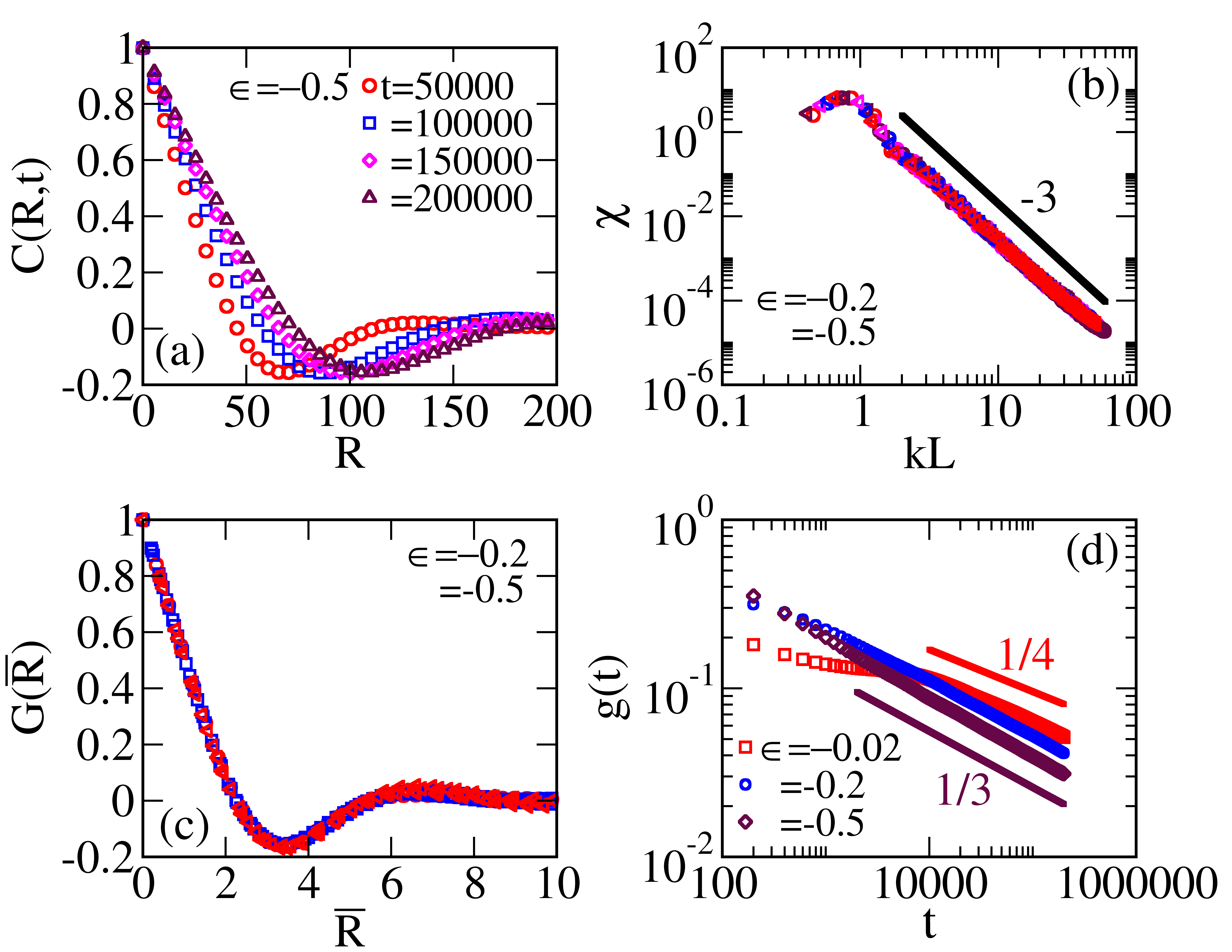}
\caption{(Color online) {\it Cahn-Hilliard  OP correlations for  various quenches :} (a) For $\epsilon =-0.5$,  data for $C(R,t)$ vs  $R$ for various times $t$; (b) Porod's Law behaviour  in  scaled structure factor $ \chi (kL)$ vs scaled wave-vector $kL(t)$,  showing data collapse in Fourier space, for all times and temperatures; (c) Dynamical scaling in $G(\bar R)$ vs  scaled separation $\bar R \equiv R/L$, showing data collapse in coordinate space, for all times, and temperatures; (d) Log-log plot of  {\it unscaled}  curvature $g(t) \equiv 1/ L(t)$ vs time $t$, showing ``guide to the eye'' indicated exponents $\alpha =1/4,1/3$, seen within the holding time $t_h$.}
\label{Fig.5}
\end{center}
\end{figure}

\begin{figure}[ht]
\begin{center}
\includegraphics[height=6cm, width=8cm]{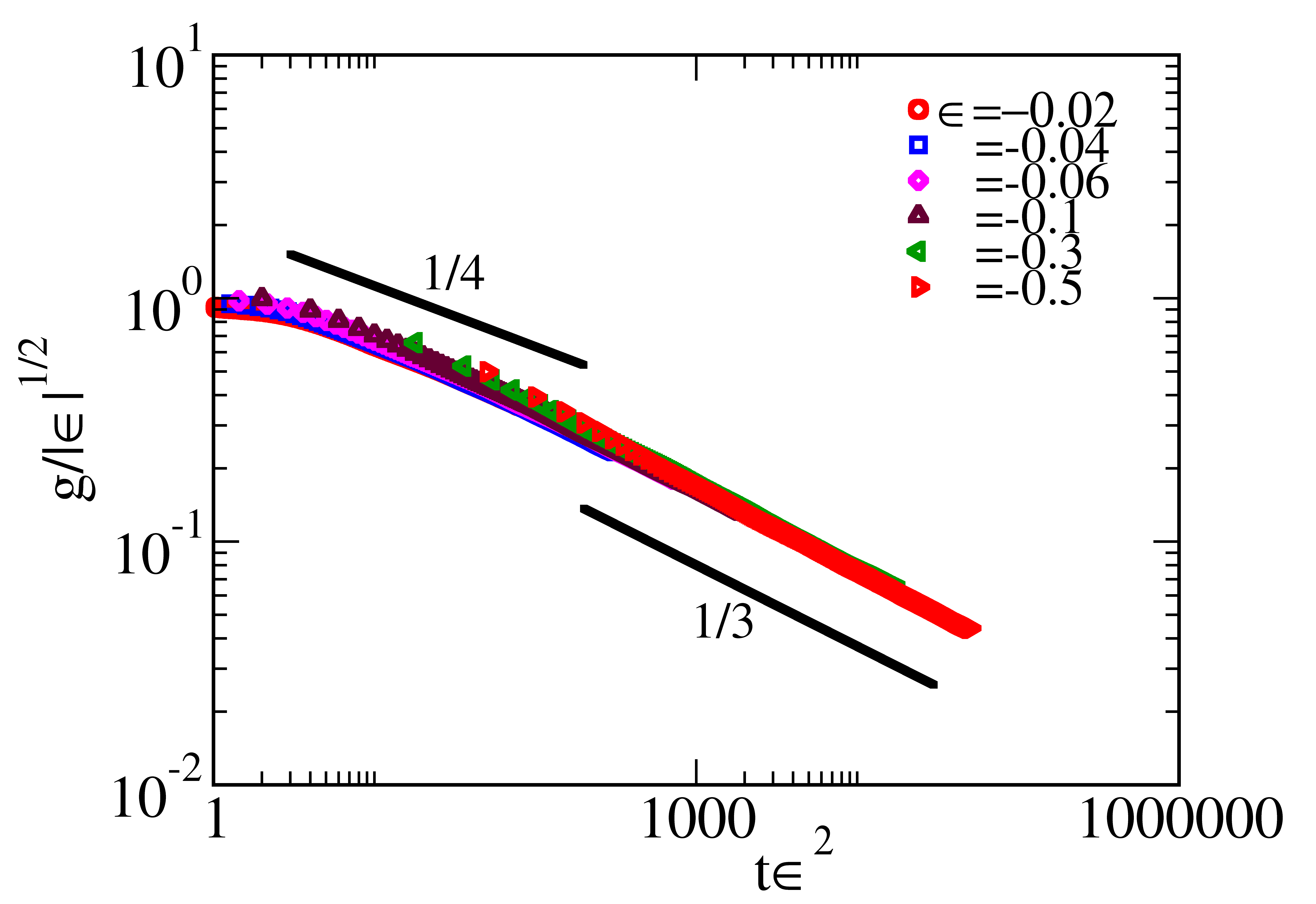}
\caption{(Color online) {\it CH  curvature coarsening  and scaled variables:} Plot of previous CH data of Fig. 5(d), in {\it scaled} variables, showing $g/{\bar \varepsilon} = g/|\epsilon|^{1/2}$ vs $t ~{\bar \varepsilon}^4 = t~ |\epsilon|^2$. There  is data collapse for all temperatures, and  guide to the eye lines indicate exponents of  $ \alpha =1/4$ for intermediate times, crossing over to  $\alpha = 1/3$ for long times. }
\label{Fig.6}
\end{center}
\end{figure}

\begin{figure}[ht]
\begin{center}
\includegraphics[height=6.5cm, width=8.5cm]{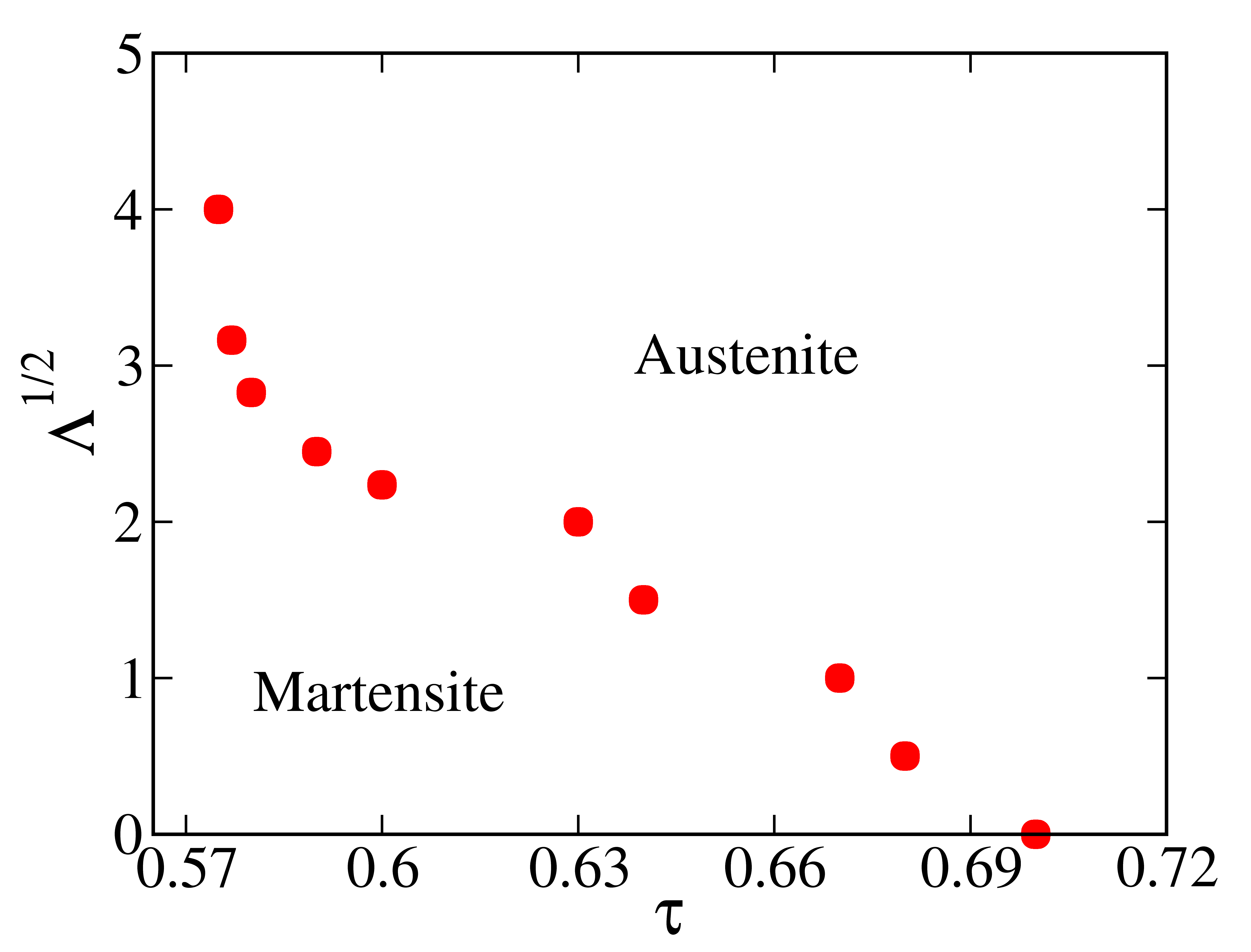}
\caption{(Color online) {\it Phase diagram:} Inverse damping  $\Lambda$ versus scaled quench temperature $\tau$, with  austenite above the transition-temperature curve, and martensite, below it. }
\label{Fig.7}
\end{center}
\end{figure}

\begin{figure}[ht]
\begin{center}
\includegraphics[height=6.5cm, width=8cm]{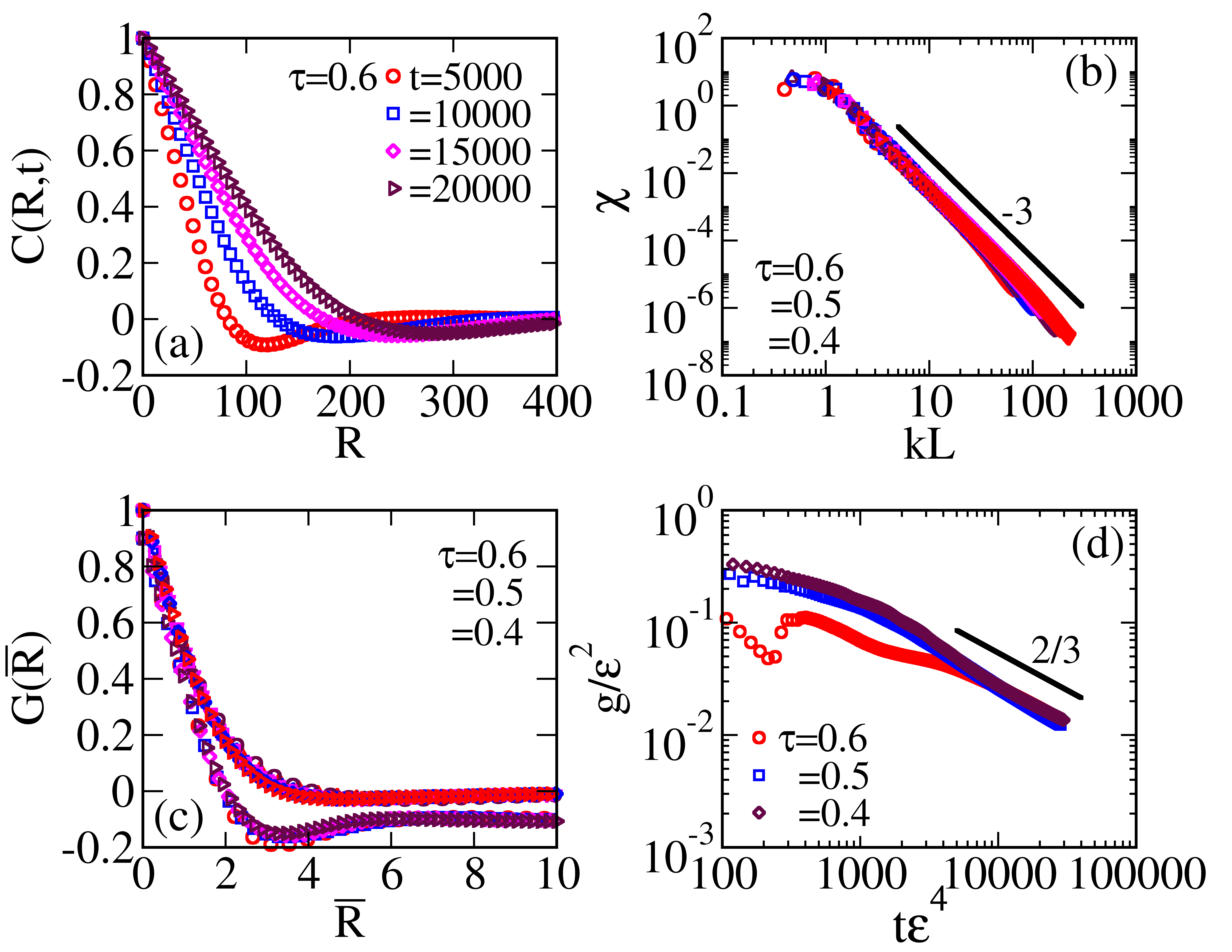}
\caption{(Color online) {\it Bales-Gooding OP correlations for  shallow  quenches:}  (a) For $\tau = 0.6$, data for $C(R,t)$ versus  $R$ for various times; (b) Porod's Law behaviour  in  scaled structure factor $\chi (kL)$ versus scaled wave-vector $kL(t)$ showing data collapse in Fourier space, for all times ; (c) Dynamical scaling in $G(\bar R)$ versus  scaled separation $\bar R \equiv R/L$ showing data collapse in coordinate space, 
for all times. The scaled curves  actually have slightly different shapes, for different temperatures, due to residual $\tau$ dependence,  and the $\tau =0.6$ curve is shifted downwards  by $0.1$, to highlight  this.For decreasing temperatures $\tau=0.5,0.4$  the curves are closer in shape and values, so the data are not relatively shifted.  (d) Log-Log plot of {\it scaled}  coarsening curvature  $g (t)/ {\bar \varepsilon} (\tau)^2$, vs $t ~{\bar \varepsilon} (\tau)^4$,  with indicated exponents $\alpha=2/3$ seen, within $t_h$. } 
\label{Fig.8}
\end{center}
\end{figure}

\begin{figure}[ht]
\begin{center}
\includegraphics[height=6.5cm, width=8cm]{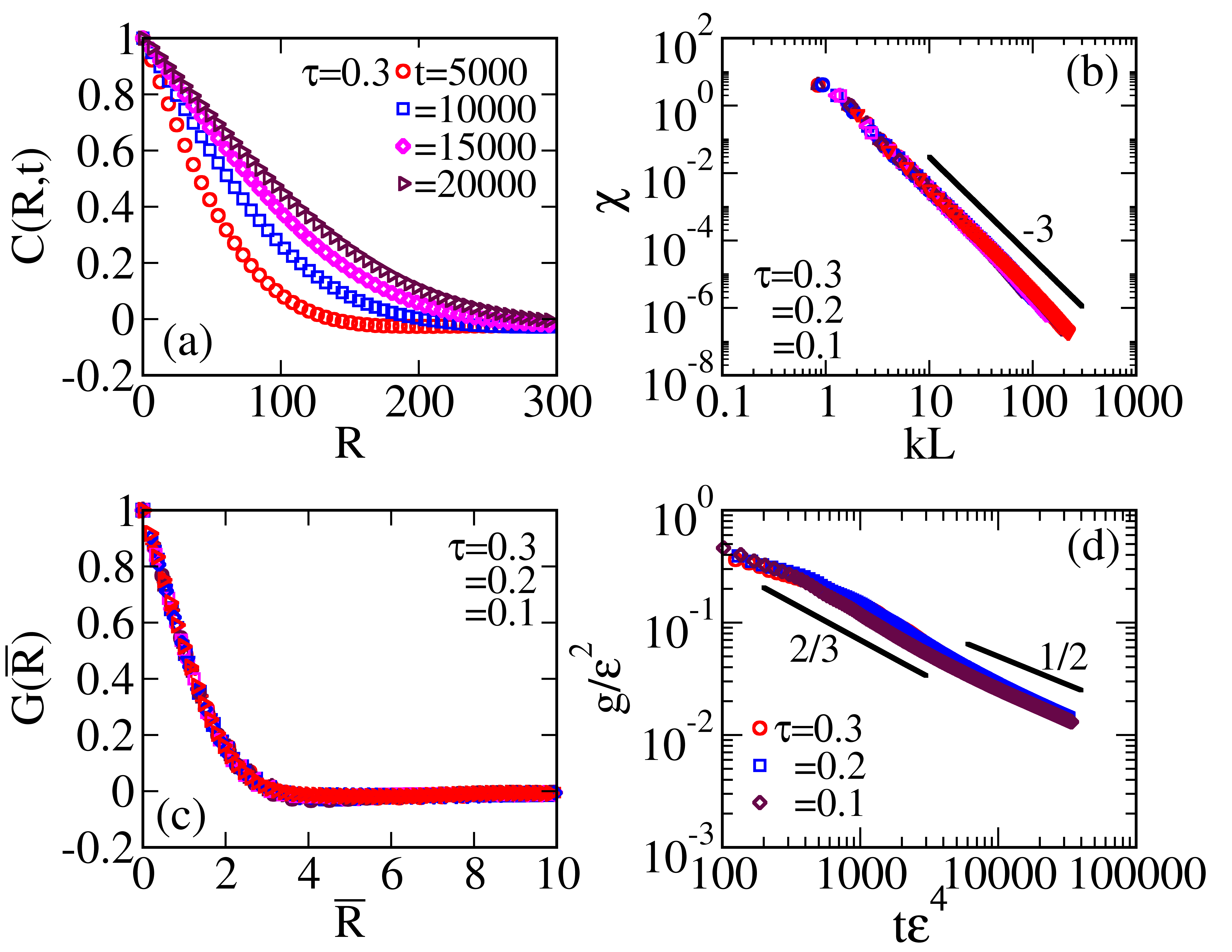}
\caption{(Color online) {\it Bales-Gooding OP correlations for moderate  quenches:}  (a) For $\tau = 0.3$,  data for $C(R,t)$ versus  $R$ for various times; (b) Porod's Law behaviour  in scaled structure factor  $\chi (kL)$ versus scaled wave-vector $kL(t)$, showing  data collapse in Fourier space for all times; (c) Dynamical scaling in $G(\bar R)$ versus scaled separation  $\bar R \equiv R/L$, showing data collapse in coordinate space, for all times, and approximately for all temperatures;  (d) Log-Log plot of scaled coarsening curvature    $g (t)/  {\bar \varepsilon} (\tau)^2$ versus   scaled time $t ~{\bar \varepsilon} (\tau)^4$, with indicated exponents$\alpha =2/3, 1/2$ seen  within the $t_h$.  } 
\label{Fig.9}
\end{center}
\end{figure}

\begin{figure}[ht]
\begin{center}
\includegraphics[height=6.5cm, width=8cm]{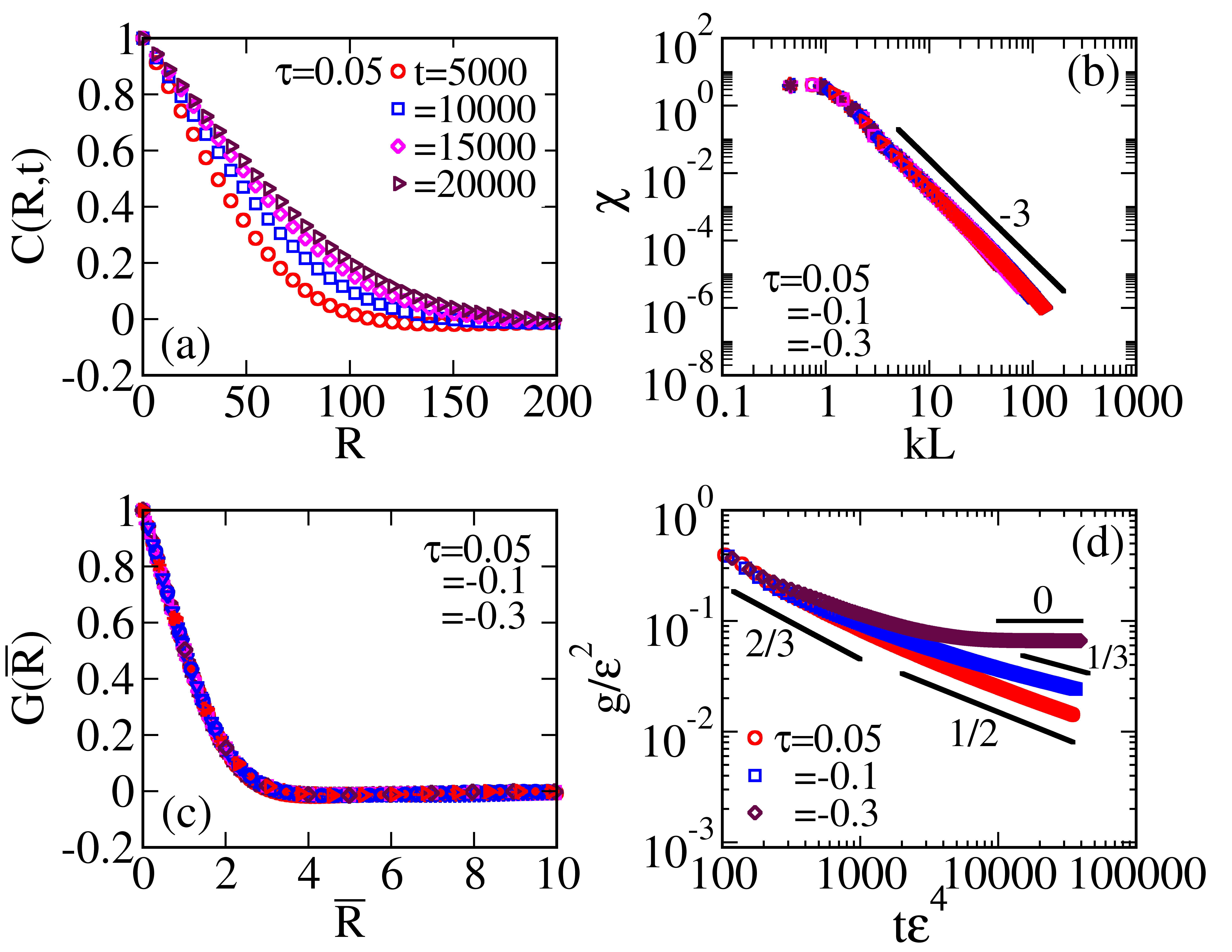}
\caption{(Color online) {\it Bales-Gooding OP correlations for deep quenches, including below the $\tau =0$ spinodal:} (a) For $\tau = 0.05$, data for $C(R,t)$ versus  $R$ for various times $t$; (b) Porod's Law behaviour  in $ \chi (kL)$ versus scaled wave-vector $kL(t)$ showing data collapse in Fourier space, for all times, and temperatures ; (c) Dynamical scaling in $G(\bar R)$ versus $\bar R \equiv R/L$ showing data collapse for all times, and approximately for all  temperatures; (d) Log-Log plot of scaled coarsening curvature $ g (t) / {\bar \varepsilon} (\tau)^2$ versus  scaled time $t ~ {\bar \varepsilon} (\tau)^4 $ with indicated exponents  $\alpha=1/2, 1/3,0$, seen within the $t_h$.   } 
\label{Fig.10}
\end{center}
\end{figure}

Figure 2(b) shows the CH case ordered-fraction $n_m (t)$ versus scaled  time $t/ t_{\rm sc} = t \epsilon^2$, where its slow rise for $T$ close to $T_c$ is due to critical slowing down near a second-order transition. As will be seen, these intermediate times can nonetheless show exponent behaviour.

Figure 3 shows snapshots of evolving relief plots of the OP $e(\vec r,t)$  versus position $(x,y)$, for early time evolutions for $t =100, 250, 1500$, and deep quenches. The strain textures or DW patterns, clearly {\it coarsen} with time. 

Figure 4 again shows snapshots of  coarsening, but now as evolving contour plots, for shallow, moderate and deep quenches.  
We will later consider several such quenches just below  $\tau= +0.6, +0.3, +0.05$, corresponding to 
$\Delta \tau (T) \equiv \tau (T) - \tau (T_4)= -0.1, -0.5, -0.8$.  
The first row is for a shallow quench, to $\tau = 0.6$, and shows, in a background of  $e=0$ austenite,  many whorl-texture droplets at early times, like a DW vapour. The second and third rows are for moderate and deep quenches  around $\tau = 0.2, -0.1$, with wandering martensite-martensite interfaces, like a DW liquid. For deep quenches, the coarsening seems to slow, or be arrested to form a DW Glass, as discussed later. 

As a benchmark, we start with the familiar CH case, in 2D and with a scalar $N_{OP} =1$ order parameter. Fig 5 shows the well-known results of dynamical scaling.

Figure 5(a) shows  $C(R,t)$ versus $R$ curves at different times for a given quench, with $\epsilon (T) = -0.5$ shown. Extracting the coarsening length and re-plotting, we find (i) data collapse of different- time curves on to a common scaling curve $C(R,t) = G(g(t) R)$, for that quench; further, (ii) data collapse of  different-quench scaling curves $G (\bar R)$ onto a {\it single} $T$-independent curve. This is consistent with (2.14), which predicted a OP-scaled CH dynamics would be independent of temperature.  The curvature exponents are consistent with the literature, with a long-time exponent \cite{R1} of $\alpha =1/3$ and, for temperatures close to $T_c$, an  intermediate-time exponent  \cite{R5} $\alpha = 1/4$. Simulations in $d =3$, show \cite{R8} the same asymptotic $\alpha = 1/3$ exponent, that is, thus independent of spatial dimension. 
 
Figure 6  shows CH case log--log plots of the scaled curvature $g/{\bar \varepsilon} =g/ |\epsilon|^{1/2}$ versus scaled time $t{\bar \varepsilon}^4 = t |\epsilon|^2$, showing data collapse, not only for all times, but also for all temperatures, as
 expected from the OP-scaled form of (2.14). Existing results \cite{R1,R8,R9} have temperature-independent exponents at  intermediate-times where $g (t)  \xi_{\rm GL} (T) >1$ as $\alpha = 1/4$;   and a long times where $g \xi_{\rm GL}(T) <1$ as the Lifshitz-Slyozov exponent of $\alpha =1/3$.  Extracting exponent values from data (as for BG case of Appendix C), we find averaged exponents and standard deviations as  
 
 $$~~~\alpha = 0.27 \pm  0.02~;~~ \alpha = 0.34 \pm 0.02,~~~~~~~(4.1) $$ 
 close to $1/4$ and $1/3$, as in previous work.
           
 Turning to the BG case quenches, Bales and Gooding in 1D find a backward-bending phase boundary of inverse damping versus temperature.
 Figure 7 shows that in the 2D case,  the phase diagram of  inverse-damping  $\sqrt {\Lambda} \sim 1/ \gamma$ versus  temperature $\tau (T)$ also has the transition occurring at lower $\tau$, for decreasing damping. Below the boundary,  the  dilute initial seeds with small $n_m$,  evolve to $n_m \rightarrow 1$ (martensite), while above it,  $n_m \rightarrow 0$ (austenite). All quenches are  to below the phase boundary. 

For $d=2$ in previous Monte Carlo simulations of a related model, a complex textural energy was parametrized by a surrogate-droplet energy that was a universal inverted parabola versus a scaled evolving textural parameter \cite{R18}.  As a check, our BG case dynamical evolutions were benchmarked against this energy parametrization  \cite{R23}.

The BG case plots of  Figs. 8-10  show numerical results analogous to the CH case, demonstrating dynamical scaling, for the three quench regimes each, just below $\tau = 0.6, 0.3, 0.05$, as mentioned.
More in detail, all the Figures show the following: 
(a) correlations $C(R,t)$ versus $R$ curves at different times, for quench to a given  $\tau (T)$; (b) scaled  structure factor  ${\chi} (kL)$, using the extracted $L(t)$,  versus $\bar k =kL$  showing Porod's tail behaviour of exponent -3 and Fourier data collapse; (c) dynamical scaling  with data collapse of different-time curves on to a common scaling curve $C(R,t) = G(g(t) R)$, for that  $\tau$ quench. However, the scaling curves $G (\bar R)$, are different for different $\tau$,  especially for the shallow quenches of Fig. 8; while for deeper quenches of Figs. 9 and 10, the curves are closer. This is consistent with the OP-scaled result of  (2.15), that shows a residual $T$-dependence  of the BG dynamics, through $\eta_{sc} (T) = \tau (T) / 3 {\bar \varepsilon} (T)^4$, that is, insensitive to $\tau$ at low temperatures.
Also they show (d) log-log plots of the coarsening curvature versus time, showing different indicated exponents in different time windows, within the holding time $t_h$. 

Figure 8 shows $\alpha = 2/3$; Fig. 9 shows $ \alpha = 2/3$ followed by $1/2$; Fig. 10 shows $\alpha = 1/2$ followed by $\alpha = 1/3$, and a peculiar flattening  to `$\alpha =0$' at low temperatures. The trapped curvature $g_0 =0.1$ is not just a finite-size effect, as for our system sizes, $g_0 \gg 1/ L_{\rm sys} \sim 10^{-4}$ .

The exponent mean values and standard deviations,  with simple arithmetic average over all temperatures, are 

$$~~~~~\alpha = 0.66 \pm 0.02 ~~~{\text and}~~~~ 0.53 \pm 0.02, ~~~~~~~~~~~~~~~~~~~~~~(4.2)$$
that are close to $\alpha =2/3$ and $\alpha =1/2$. See Appendix C for a time-window procedure for extracting exponents.

\begin{figure}[ht]
\begin{center}
\includegraphics[height=3.5cm,width=8.5cm]{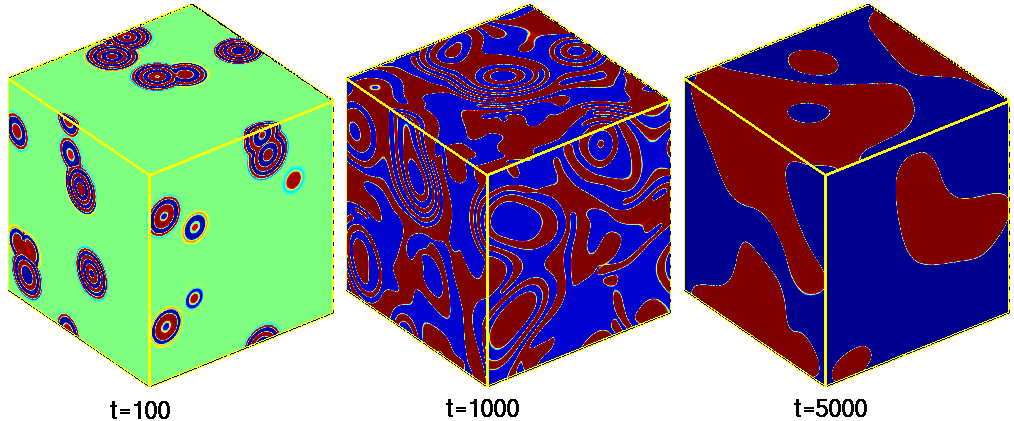}
\caption{(Color online) {\it Contour plots of 3D strain evolutions:} Snapshots for various times show the OP strain $e ({\vec r}, t)$ 
evolving after a quench, in a system of size $512^3$.}
\label{Fig.11}
\end{center}
\end{figure}

\begin{figure}[ht]
\begin{center}
\includegraphics[height=6.5cm, width=8cm]{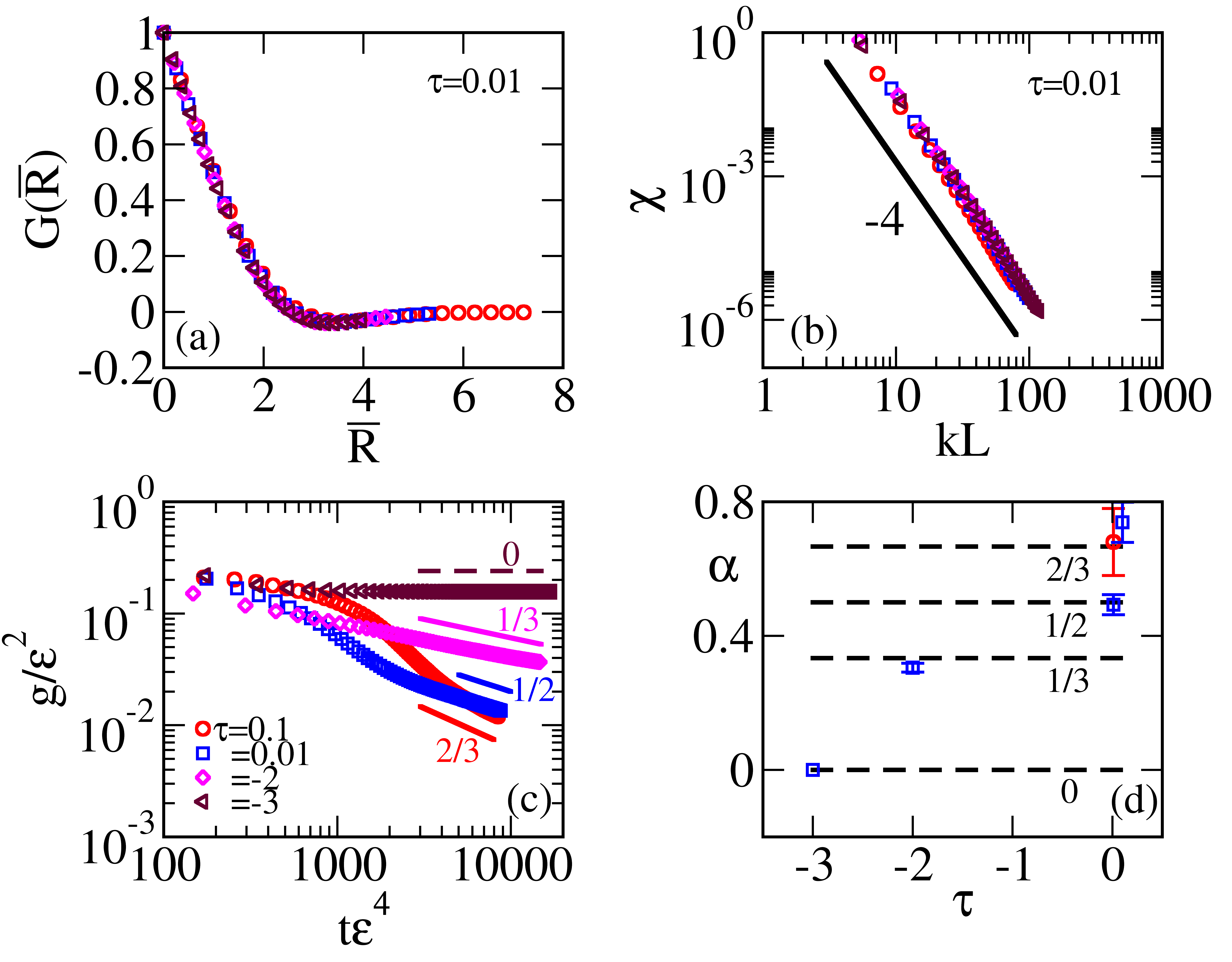}
\caption{(Color online) {\it Bales-Gooding OP correlations in 3D for $\tau =0.01$ :}  (a) Dynamical scaling in $G(\bar R)$ versus $\bar R \equiv R/L$ for various times $t$, showing data collapse in coordinate space;  (b) Porod's Law behaviour  in $ \chi (kL)$ versus $kL$ showing data collapse in Fourier space, with slope $-(d + N_{\rm OP}) = -4$; (c) Log-Log plot of scaled coarsening curvature $ g (t) / {\bar \varepsilon}^{2}$ versus scaled time $t \bar \varepsilon^4$ showing various indicated exponents; (d) Fitted exponent values $\alpha$ versus quench temperature $\tau$, showing that  the 3D values close to $2/3,1/2,1/3,0$ are the same as in 2D.   } 
\label{Fig.12}
\end{center}
\end{figure}

All this is for 2D. We have also considered 3D coarsening textures, as  shown in surface contour plots of Fig. 11. The 3D case also shows dynamical scaling as shown in Fig. 12, with the Porod's law exponent now $-(d+1) =-4$, and data collapse as before. For successively deeper quenches, the exponents are found to be again close to $\alpha =2/3, 1/2, 1/3, 0$, so the coarsening exponents are independent of spatial dimension.

\section{Correlation-function  Dynamics and Dynamical Scaling}

 The exponent behaviour has been extracted  from various OP dynamics by many authors \cite{R1,R5,R6,R11}, including  Siggia through insightful heuristic arguments; Bray and Rutenberg through matching global and local dissipation;  Ohta, Jasnow, and Kawasaki; Bray, and Puri, and Mazenko, through a fluctuating domain wall approach; and Langer {\it et al}. through a self-consistent truncation of correlation function dynamics.
 We suggest a complementary, and systematic { \it curvature kinetics} approach, to extract exponent behaviour.
 
In our theoretical analysis, we will use throughout, the OP-scaled form of the CH and BG dynamics, that fully or mostly, scales out the OP  temperature dependence.
We first obtain the dynamics of the  two-point OP-OP correlation as done  previously, elsewhere \cite{R5}. 

\subsection{Evolution of correlation function}

The correlation function dynamics can be derived  \cite{R5, R24}. from a given order parameter dynamics, such as the OP-scaled  dynamics of (2.14) and (2.15).
From (2.14), the ``mass-conserving'' Cahn-Hilliard correlation function dynamics is
$$
 \frac{\partial}{\partial t} C(R,t)  = {\vec \nabla_{\vec R}}^2 C^{(\mu)}(R,t). ~~~(5.1)
$$

From  (2.15) we similarly get a ``momentum-conserving'' Bales-Gooding  correlation dynamics,

$$
\Lambda \frac{\partial^2}{\partial t^2} C(R,t)= {\vec \nabla_{\vec R}}^2 \left[~  C^{(\mu)}(R,t) + \frac{\partial C(R,t)}{\partial t} \right].~~~(5.2)
$$

The chemical potential-order parameter correlation or $\mu-{\text OP}$ correlation of (3.3),  has  separate Landau and Ginzburg contributions, 
$$
C^{(\mu)}(R,t) \equiv {C}^{(\mu_L)}(R,t) + {C}^{(\mu_G)} (R,t) , ~~~ (5.3)
$$
where the Ginzburg term depends directly on the OP-OP correlation, 
$$
{C}^{(\mu_G)} (R,t) \equiv \left\langle \left[\frac{1}{2} \frac{\partial f_G }{\partial e (\vec r, t)} \right] e (\vec {r^{\prime}}, t) \right\rangle $$
$$= -{\vec \nabla_{\vec R}}^2 C (R, t),  ~~~(5.4a)
$$
while the Landau contribution carries the higher powers of the strain 
$$
{C}^{(\mu_L)} (R,t) \equiv  \left \langle \left[\frac{1}{2} \frac{\partial f_L}{\partial e (\vec r, t)}\right] e (\vec {r^{\prime}}, t) \right\rangle
$$
$$
\equiv  -\left \langle f_0 (e(\vec r,t))  e(\vec r,t) e(\vec {r^{\prime}} , t) \right \rangle, ~~~(5.4b)
$$
where the BG-case scaled polynomial factors $f_0 (e)$ have been given earlier. 

\subsection{Dynamic scaling ansatz}

We assume the correlation functions have a dynamic scaling form, 
and  insert this as an {\it ansatz} or trial solution,
 
 $$
 C(R,t) = G (\bar R), ~~(5.5a)$$
where the argument of the scaling function is henceforth
  $$
  {\bar R} \equiv g(t) R. ~~(5.5b)
  $$
   
 For the  Landau part of  $\mu-$OP correlation function with three lengths $\xi_0, R, L$,we can similarly assume a general scaling form in  ratios $R/L(t), \xi_0/ L(t)$:
 
$$ 
 {C}^{(\mu_L)} (R, t) =( g \xi_0 )^{\delta_L}   G^{(\mu_L)} (\bar R, g \xi_0), ~(5.6)
$$
where $G^{(\mu_L)}$ is some scaling function, and $\delta_L$ some universal exponent.

The model dynamics can be written in terms of ${\bar R}$ derivatives of the scaling functions, and time-derivatives of the curvature  $g(t)$. 

The Laplacians are derivatives in $\bar R$, with no angular derivative surviving, when acting on isotropic functions
 $$
 \vec \nabla_{\vec R}^2 = g^2 {\hat D} ~; {\hat  D} \equiv  \left [ \frac{\partial^2}{\partial{\bar R}^2} +\frac{ (d-1)}{\bar R} \frac{\partial}{ \partial {\bar R}} \right ] .~(5.7)
 $$

The time derivatives are
$$
 \frac{\partial G(g R)}{\partial t}=\left\{\frac{\dot g}{g}\right\} {\bar R} G^{\prime}({\bar R});  ~~~(5.8a)
$$

 $$
  \frac{\partial^2 G( g R)}{\partial t^2}= \left\{\frac{\ddot g}{g}\right\}{\bar R} G^{\prime}(\bar R) + \left\{\frac{\dot g}{g}\right\}^2{\bar R}^2 G^{\prime\prime}(\bar R),~~(5.8b)
$$
where  primes denote derivatives $G' \equiv d G/ d{\bar R}, G^{\prime\prime} \equiv  d^2 G/ d{\bar R}^2$ and so on.
Note the prefactors of the curvature time-derivatives  contain $G^{\prime} (\bar R)$, which is slowly varying at its turning points $G^{\prime\prime}(\bar R=\bar R_0) =0$.

The Cahn-Hilliard (CH) dynamics is, setting $\xi_0 =1$,
$$\left\{\frac{\dot g}{g}\right\} {\bar R} G^{\prime}({\bar R}) =  g^{ \delta_L} \left[g^2 {\hat D} {G}^{(\mu_L)}  ({\bar R}, g)\right]   - g^4 \left[ {\hat D}\right] ^2 G~.~~(5.9a) $$

The Bales-Gooding (BG) dynamics is

$$\Lambda \left[\left\{\frac{\ddot g}{g}\right\}{\bar R} G^{\prime}(\bar R) + \left\{\frac{\dot g}{g}\right\}^2{\bar R}^2 G^{\prime\prime}(\bar R)\right] =$$
$$ g^{2+ \delta_{L}}\left[ {\hat D} G^{(\mu_{L})}({\bar R})\right]-g^4 \left[{\hat D}^2  G\right]  + g^{2} \left\{ \frac{\dot g}{g} \right\} {\hat D} ({\bar R} G^{\prime}) .~~(5.9b)$$

 The derivative operators  of (5.7) yield 
$$ {\hat D} G^{(\mu_L)}  =\frac{1}{{\bar R}^2} \left[ (d-1) {\bar R} {G^{(\mu_L)}}^{\prime}  + {\bar R}^2 {G^{(\mu_L)}}^{\prime\prime} \right ]~; ~~ (5.10a)$$

$$  {\hat D}^2  G = \frac{1}{{\bar R}^4} [(3-d)(d-1) ( {\bar R} G^{\prime} - {\bar R}^2 G^{\prime\prime}) $$
$$ +2(d-1){{\bar R}^3}G^{\prime\prime\prime} + {\bar R}^{4} G^{\prime\prime\prime\prime} ]~;~~~~~~(5.10b)$$

$${\hat D} \left[ {\bar R} G^{\prime} \right]  =  \frac{1}{{\bar R}^2} \left[ (d-1){\bar R} G^{\prime} + (d+1){\bar R}^2 G^{\prime\prime} + {\bar R}^3 G^{\prime\prime\prime} \right].~(5.10c)$$

Collecting terms above, and dividing through by  ${\bar R} G^{'}({\bar R})~$,  we have
for the CH case,
$$
 \frac{-\dot g}{g}  = J_3 (\bar R) g^{2 + \delta_L} + J_4 (\bar R) g^4;~~(5.11a)
$$

and  for the BG case,
$$
 -\Lambda \left[ \frac{\ddot g}{g} + I_2 (\bar R) \left\{\frac{\dot g}{g}\right\}^2 \right]+  K_1 (\bar R) \left\{\frac{-\dot g}{g}\right\} g^2 $$
 $$ = J_3 (\bar R) g^{2 +\delta_L} + J_4 (\bar R) g^4 . ~~(5.11b)
$$
Here  $J_n$ are  $\bar R$-dependent  coefficients of  powers  $g^n$ of the force terms, while $K_1 (\bar R)$ is a coefficient of  the kinetic  damping term.  (The Landau term   coefficient is called $ J_3 (\bar R)$, as $\delta_L =1$, later.)

The coefficients in  (5.11)   are
  $$K_1 (\bar R) \equiv  \left[ \frac{-1}{{{\bar R}} ^2}\right]  \left[ (d-1)+ (d+1) I_2  + I_3 \right] ~;~~(5.12a)$$

$$J_3 (\bar R, T) \equiv  \left[\frac{-1}{ {\bar R} ^2}\right] [ (d-1){\bar R}{G^{(\mu_L)}}^{\prime} +{\bar R}^2 {G^{(\mu_L)}}^{\prime\prime}] /{\bar R} G^{\prime}~;(5.12b)$$

$$J_4 (\bar R) \equiv \left[\frac{1}{ {\bar R} ^4}\right] \left[(3 -d) (d -1) (1- I_2)
   +2 (d-1) I_3  + I_4  \right], ~(5.12c)$$
with all in terms of the derivative ratios of $G(\bar R)$,
 
 $$ I_2 \equiv \frac{ {{\bar R}}^2  {G^{(\mu_L)}}^{\prime\prime}       ({\bar R})}{ {{\bar R}} G^{\prime}({\bar R})}~;$$

 $$~I_3 (\bar R) \equiv \frac{ {{\bar R}}^3 G^{\prime\prime\prime}({\bar R})}{ {{\bar R}} G^{\prime}({\bar R})}~;
I_4 (\bar R) \equiv \frac{ {{\bar R}}^4 G^{\prime\prime\prime\prime}({\bar R})}{ {{\bar R}} G^{\prime}({\bar R})}~.~~(5.13)$$

So far, this is formally exact, with the only input being a dynamical-scaling trial solution. 
The coefficient  $ J_3$ contains the yet unspecified $G^{(\mu_L)}$,
that carries the higher-order correlations in the OP. Its further evolution equations would induce a dynamically coupled, infinite hierarchy \cite{R5} . A {\it closure approximation} is needed, and there must be a  {\it coefficient evaluation}  at some physically  motivated,  expanding-front value of  $\bar R$. 
 
\subsection{ Approximations}
 
 {\it a) Closure Approximation}: \\

The correlation between the Landau chemical potential and the OP is

$$ C^{(\mu_L)}= -\left\langle f_{0}(e(\vec r,t)) ~e(\vec r,t) e (\vec {r^{\prime}},t) \right\rangle. ~~(5.14a)$$
and the {\it f}actor $f_0 (e)$ carries higher order powers of $e$, that induce the correlation hierarchy.  
For a {\it uniform} or bulk order parameter $e =\pm 1$, the Landau part of the chemical potential vanishes, $\mu_L \sim \partial f_L / \partial e =- e f_0 (e) =0$. The correlation 
${C}^{(\mu_L)} (R, t)$ has contributions only from the {\it non}-uniform  OP regions around domain walls, where $f_0 (e) \neq 0$  over a thickness $\xi_0$ between the competing bulk values.  Thus  ${C}^{(\mu_L)}$ is a  correlation between an OP and many possible DW. It  decreases for decreasing  DW thickness $\xi_0$, and     $ C \sim (g \xi_0)^{\delta_L}$ as in (5.6).  Interpreting the scaling function $G^{(\mu_L)}$  as a correlation between the OP and a single DW  the prefactor is then the probability of finding a DW, enabling an estimate of $\delta_L$.  

In a  coarsening volume $L (t) ^2$, the probability of finding a scalar-OP DW is  roughly $(\xi_0 L)/ L^2 = \xi_0/ L = \xi_0 g$.  The exponent in (5.6) for $N_{\rm OP} =1$ is   then $\delta_L = 1$. 

As Bray \cite{R1} has noted, for general number of OP components, and in $d$ spatial dimensions, the vanishing at a DW of  {\it all} the $N_{\rm OP}$ components, corresponds to a `surface' of reduced dimension  $d-N_{\rm OP} >0$.  Thus more generally,  in a  coarsening volume $L^d$, the probability of finding a DW is  roughly  $(\xi_0^{N_{\rm OP}} L ^{d-N_{\rm OP}}) / L^d = (\xi_0/ L)^{N_{\rm OP} }= (\xi_0 g)^{N_{\rm OP}}$. The exponent  for general $N_{\rm OP}$ is   then  $\delta_L = N_{\rm OP}$, 
and (5.6) is taken as

$$
{C}^{(\mu_L)} (R, t) =  (g \xi_0) ^{N_{\rm OP}} G^{(\mu_L)} (\bar R, g \xi_0). ~~~ (5.14b)
$$

We make a simple closure approximation for the domain-wall scaling function $G^{(\mu_L)} (\bar R, g\xi_0)$. To leading order in $g \xi_0$, we take $G^{(\mu_L)}(\bar R, g \xi_0) \simeq G^{(\mu_L)} (\bar R, 0)$. (However in Section VI, we will consider possible higher curvature corrections in $g \xi_0$ from $G^{(\mu_L)}(\bar R, g \xi_0)$, in a toy model for coarsening arrest.) We then replace  the DW factor $f_0 (e)$ by  its  spatial average  ${\bar f}_0 (T) = \langle f_0 (e(\vec r)) \rangle$. This yields, as in Appendix A, 

$$G^{(\mu_L)} (\bar R, g \xi_0)  \simeq  G^{(\mu_L)} (\bar R, 0) = - {\bar f}_0 (T) G(\bar R).~~(5.14c)  $$
 Thus reasonably, $G^{(\mu_L)}$ has the same $\bar R$ dependence as $G$, as  also holds in other  approximations \cite{R5}.
 
Inserting this closure approximation into the Landau term coefficient of (5.12b),
 
$$~J_3 (\bar R) \simeq \frac{{\bar f}_0 (T) } { {\bar R} ^2} \left [(d-1) + I_2 (\bar R)\right]~. ~(5.15)$$

{\it b) Coefficient evaluations}\\

The correlation-function dynamics  of (5.11) is now closed, but has a peculiar form, of an equation nonlinear in the curvature  $g  (t)$ and its time derivatives; with coefficients linear in $G (\bar R)$ and its scaled space derivatives.

The coefficients are evaluated at some constant separation $\bar R$.
As the slope  $G'(\bar R)$ is a prefactor in  the kinetic terms,  it is natural to focus on where it is slowly varying,  at its own  turning point,                                                
 $G''({\bar R}_0)  =0$. This defines a dominant curvature front $R = {\bar R}_0 L(t)$. For both CH and BG dynamics, $G(\bar R)$ first has a minimum ($G'' >0$), and then a maximum ($G'' < 0$), so this point where $G'' =0$ is somewhere in between.  We evaluate all coefficients at  the first turning point  $\bar R ={\bar R}_0$ of $G'(\bar R)$, that is also a non-stationary {\it inflection} point of the scaled correlation $G(\bar R)$. 
 
 With  $G'' ({\bar R}_0) =0$ and hence $I_2 ({\bar R}_0)= 0$,  the coefficients are

$$K_{10}\equiv  K_1 ({\bar R}_0)=  \left[ \frac{-1}{{{\bar R}_0} ^2}\right]  \left[ I_3 (\bar R_0) + (d-1)\right] ~; ~~(5.16a)$$

$$J_{30}\equiv J_3 ({\bar R}_0)  = \frac{{\bar f}_0 (T) ~(d-1)}{ {\bar R_0}^2 } ~;~~~(5.16b)$$

$$J_{40} \equiv  J_4({\bar R}_0) $$
$$ = \left[\frac{1}{ {\bar R_0} ^4}\right]  \left[ I_4 (\bar R_0)   +2 (d-1) I_3 (\bar R_0)  +(3 -d) (d -1) \right].~(5.16c)$$
Note that  $I_2=0$  in the inertial term of (5.8b), suppresses nonlinearities, leaving 
just a curvature acceleration, $\sim \Lambda \ddot{g} / g$.
Fits to $G(\bar R)$ in  Appendix B yield   ${\bar R}_0 >1$. 

An alternative, and equivalent choice for coefficient evaluation is  where the OP gradient-gradient correlation, or effective DW-DW correlation 

$$\Gamma (\bar R) \equiv g^{-2} \langle \nabla_{\vec r}~ e(\vec r,t). \nabla_{\vec {r^{\prime}}}~ e(\vec {r^{\prime}},t) \rangle ~(5.17)$$ 
 flattens to zero.  Fig 15d of Appendix B shows that this flattening occurs near the previous  choice, of the first inflection point  ${\bar R}/ {\bar R}_0 =1$. This gives a physical justification to our evaluation choice. 
 
 The simple approximations made here are solely for the limited purpose of determining the now-constant coefficients, of a curvature kinetics.

 \section{Curvature Kinetics}

The dynamics is now in terms of $g(t)$ only, and  can yield exponent behaviour
for appropriate coefficient signs; with possible crossovers in time between these exponents. 

The  curvature kinetics,   derived from a given order parameter dynamics,  yields five main results.  \\
i) There are time regimes where the curvatures decay as single power laws in time $g \sim 1/t^\alpha$.\\
ii) The exponents  $\alpha$ are ratios of integers, induced directly from the  integer powers of the curvatures, in each derived kinetics.\\
iii) In addition to the long-time exponents, there can also be different exponent behaviour at intermediate times, from two different force terms sequentially balancing the kinetic term. \\
iv) The exponents are  manifestly independent of spatial dimension $d$ that can be scaled out, but can depend on the number of order parameter components $N_{\rm OP}$.\\
v) The scaled kinetics can be solved analytically in some cases, providing a universal scaling function of  curvature versus time.\\ 

The curvature kinetics for the CH case is
$$
 \frac{-\dot g}{g}  = J_{30} g^{2 + N_{\rm OP}} + J_{40} g^4.~~(6.1)
$$

The  curvature kinetics for the BG case is,
$$
 -\Lambda \left[ \frac{\ddot g}{g} \right]+  K_{10} \left\{\frac{-\dot g}{g} \right\} g^2 $$
 $$ = J_{30} g^{2 +N_{\rm OP}} + J_{40} g^4 . ~~(6.2)
$$

We now scale times  and curvatures in crossover values  $t_{\rm cr}, g_{\rm cr}$, and define
$$ \bar t \equiv  t/ t_{\rm cr}~; ~~ \bar g \equiv g/ g_{\rm cr}. ~~(6.3)$$

The 'dot' notation henceforth is $\dot X \equiv dX / d{\bar t}$,  and we pull out  the coefficient signs  $\sigma_n$ through  $J_{n0} = \sigma_n |J_{n0}|$.

 The  curvature kinetics for the CH case is 
$$
 - {\dot {\bar g}}/ {\bar g}= \sigma_3 \left\{t_{\rm cr} g_{\rm cr}^{2+N_{\rm OP}} |J_{30}|\right\}  {\bar g}^{2+N_{\rm OP}} + \sigma_4  \left\{t_{\rm cr} g_{\rm cr}^4 |J_{40}| \right\} {\bar g}^4 . ~(6.4a)
$$

Choosing  both the curly brackets to be unity,
$$  g_{\rm cr} = \left[\frac{|J_{30}|}{ |J_{40}|}\right]^{1/\lambda}; ~~
   t_{\rm cr} =\left[ \frac{|J_{40}|^{2+N_{\rm OP}}}{~|J_{30}|^4}\right]^{1/\lambda}~~ (6.4b). 
$$
where $\lambda =  2-N_{\rm OP}$. For the special case $N_{\rm OP} =2$, there is a line of possible scalings, $t_{\rm cr} {{\bar g}_{\rm cr}} ^4 = 1/(|J_{30}| +|J_{40}|)$.

As discussed in Appendix B, we find from the CH case fits to the data, that  $\bar R_0 \sim 4.4$,   independent of $\tau$, and  $J_{30}, J_{40}$ are positive, or $\sigma_3 = \sigma_4 =1$. The CH scaled curvature kinetics is then

  $$ - {\dot {\bar g}}/ {\bar g} =    {\bar g}^3 +   {\bar g}^4 . ~(6.4c)\ $$

  The curvature kinetics for the Bales-Gooding case is

$$
- \left\{\Lambda / t_{\rm cr} ^2\right\}  {\ddot {\bar g}}/ {\bar g}+    \sigma_1 \left\{|K_{10}| g_{\rm cr} ^2/ t_{\rm cr}\right\} {\bar g}^2 (- {\dot {\bar g}}/ {\bar g})
 $$
 $$ =\sigma_3 \left\{ |J_{30}| g_{\rm cr}^{2+N_{\rm OP}} \right\}  {\bar g}^{2+N_{\rm OP}} + \sigma_4 \left\{|J_{40}| g_{\rm cr}^4 \right\} {\bar  g}^4. ~~(6.5a)
$$

Dividing through by $\{|K_{10}| g_{\rm cr} ^2/ t_{\rm cr}\}$, and choosing the  crossover scales such that the resultant  prefactors are unity as before,

$$ g_{\rm cr} =\left[\frac{ |J_{30}|}{ |J_{40}|}\right]^{1/\lambda}; ~~ t_{\rm cr} = |K_{10}| ~\left[\frac{ |J_{40}|^{N_{\rm OP}}}{  ~ |J_{30}|^2}\right]^{1/\lambda};$$
$$~ {\Lambda}^{\prime} \equiv \frac{\Lambda}{|K_{10}| g_{\rm cr} ^2 t_{\rm cr}} = \frac{\Lambda  |J_{40}|}{ |K_{10}|^2}. ~(6.5b)$$
where $\Lambda^{\prime}$ is independent of $N_{\rm OP}$.

 For the BG case, $\bar R_0$ and hence the coefficients, depend on $\tau$, as in Fig 15c of Appendix B. While  $K_{10} , J_{40}$ are positive, or $\sigma_1 =\sigma_4 =+1$,  the  sign $\sigma_3$ of $J_{30}$ is  that of $\bar f_0 \sim(1- \tau / \tau_f)$, as in Appendix B.
The BG scaled curvature kinetics is then
$$
-{ \Lambda} '  {\ddot {\bar g}}/ {\bar g}  +    {\bar g}^2 (- {\dot {\bar g}}/ {\bar g}) = \sigma_3   {\bar g}^3 +  {\bar  g}^4. ~~(6.5c)$$

Note that in both the CH and BG cases,  $d$ only enters the coefficients, and can be scaled out. The exponents are then predicted to be                                                                       {\it independent of spatial dimension}, as is indeed found in simulations, for the CH case \cite{R1,R9}, and in the BG case of Fig 12. 

We now turn to power-law solutions and their regimes.  

\subsection{ Exponent regimes for CH equation}

For  a pure power-law decay, $\bar g(t) =  {\bar g}_\alpha  /\bar t^\alpha$, the  time derivative terms in (5.11)  are independent of the prefactor $\bar g_\alpha$; and the time powers are independent of $\alpha$:
$$
 \frac{\dot{\bar g}}{\bar g}=\frac{-\alpha}{\bar t}; ~~\frac{\ddot {\bar g}}{\bar g}= \frac{\alpha(1+\alpha)}{{\bar t}^2}.~(6.6)
$$

For asymptotic vanishing of the curvature  $\bar g(t) \rightarrow 0$, the balancing of kinetic terms with the lowest power of $g < 1$ determines the long-time behaviour; while a balancing with higher powers of curvature determines the intermediate-time behaviour.

With $\sigma_3 = \sigma_4 = +1$, as in Appendix B, the CH curvature kinetics is $(-{\dot {\bar g}}/{\bar g}) = {\bar g}^3 + {\bar g}^4$.
The  kinetic term can balance the two forces sequentially, resulting in two exponents: $(-\dot{\bar g}/{\bar g}) = {\bar g}^n$, with $n =3,4$, with power-law solutions  ${\bar g} = {\bar g}_\alpha / {\bar t}^\alpha$, with $\alpha = 1/n$ and $\bar g _\alpha = (1/n)^{1/n}$. 

In previous results \cite{R9},  from heuristic arguments, the  ${\bar t}^{1/4}$-regime is associated with diffusion of material along interfaces, while the ${\bar t}^{1/3}$-regime is associated with bulk diffusion. In the curvature kinetics approach, these  physical results are derived directly,  yielding  the $1/4$ exponent from the Ginzburg term, and the $1/3$ exponent from the Landau term. 

We go back to the scaled CH dynamics of (6.4c)  and note it can be integrated exactly to yield a theoretical scaling function. For $N_{\rm OP} =1$,

$${\bar t}  =   I(1/{\bar g}) ~~~~~~~~~~~~~~(6.7a)$$
where
$$I(Y) =  [\sum_{\ell = 1,2,3}  \{ (-1)^{\ell +1} Y^\ell / \ell\}  - \ln |1+ Y| ] ~~ (6.7b)$$
where the sum is the first three terms of an expansion of the logarithm $\ln (1+(1/{\bar g}))$.
For $Y \ll 1$, the leading term is $Y^4$, yielding $\bar g \sim1/ {\bar t}^{1/4}$, while for $Y \gg 1$ the leading term is $Y^3$, yielding $\bar g \sim  1/{\bar t}^{1/3}$.

For multicomponent \cite{R1,R10}  or 'vector' OP with $N_{\rm OP} \geq 2$,  the Landau term ${\bar g}^{2+ N_{\rm OP}}$ is comparable to the Ginzburg term ${\bar g}^4$ for $N_{\rm OP} =2$; and smaller than it, for $N_{\rm OP} > 2$. Hence  the long-time exponent is predicted to be $\alpha =1/4$ for vector order parameters. This is again consistent with known  2D simulation results  \cite{R1}, that  yield a long-time  falloff  of ${\bar g} \sim 1/(t \ln t)^{1/4}$ for $N_{\rm OP} =2$, and of $\sim 1/ t^{1/4}$ for $N_{\rm OP} > 2$. The intermediate time exponents are predicted to be $\alpha = 1/ (2 + N_{\rm OP})$, or $ 1/5, 1/6..$ for $N_{\rm OP} =3, 4..$ .

\subsection{ Exponent regimes for BG equation}  

From Appendix B, the coefficient $J_{30} \sim \bar f_0(T) \sim -(1 - \tau / \tau_f)$ goes from negative to positive on cooling through $\tau = \tau_f \sim +0.3$. Simulations further show there is a  possible  flattening of the curvature for quenches below some $\tau_g \sim -0.3$.
  Hence we consider three temperature quench ranges, with characteristic exponents.

\subsubsection{$ \tau > \tau_f > \tau_g$  quench range} 

Here $\sigma_3 = -1$, and the scaled curvature kinetics is 

$$
-{ \Lambda} ' ~ {\ddot {\bar g}}/ {\bar g}  + {\bar g}^2 (- {\dot {\bar g}}/ {\bar g}) = -{\bar g}^{2+N_{\rm OP}} + {\bar  g}^4. ~~(6.8)$$

For scalar order parameters $N_{\rm OP} =1$, a balance between the Landau term $ {\bar g}^3$ and the acceleration term $\Lambda/ {\bar t}^2$ yields $\bar g \sim 1/ {\bar t}^{2/3}$.  A  balance between the Ginzburg term ${\bar g}^4$ and the damping term ${\bar g}^2 /{\bar t}$ supports $\bar g \sim 1/ {\bar t}^{1/2}$. Since damping should dominate  acceleration at late times, the kinetics predicts $\alpha = 2/3$ in the acceleration-dominated  or inertial regime at intermediate times; and $\alpha = 1/2$ in the damping-dominated regime at long times. This explains the behaviour of Figs. 8-10. Simulations in other models with inertial terms, \cite{R10}, can show  other exponent sequences of $\alpha =1, 1/2$.

For vector order parameters with $N_{\rm OP} > 2$, the  acceleration dominated intermediate-time regime shows exponents $\alpha = 2/(2 +N_{\rm OP}) =2/5,1/3..$ for $N_{\rm OP} = 3,4..$;  while the damping dominated late-time regime shows  exponent $\alpha =1/2$ as before. For the special case $N_{\rm OP} =2$,  one needs higher order curvature terms, similar to the toy model as given below, that could yield $\alpha = 2/5$, in this quench range.

\subsubsection{$ \tau_f  \geq \tau >\tau_g$  quench range}  

Here $\sigma_3 =+1$, and  the Landau term is the wrong sign to balance the acceleration. In fact, going back to the unscaled kinetics if $\tau = \tau_f$, then $J_3 =0$, and only the Ginzburg term survives, to balance both the acceleration and damping. Inserting  a pure power-law solution
${\bar g} = {\bar g}_{\alpha}/ {\bar t}^{\alpha}$,  

$$
{ {\bar  g}_\alpha}^4 /{\bar t}^{4\alpha}  -    ( {{\bar g}_ \alpha}^2 / {\bar t}^{1+2\alpha})  +({ \Lambda}^{\prime}  \alpha (\alpha +1) /{\bar t}^2) \simeq 0, ~~ (6.9)$$
This gives  $\alpha = 1/2$, with a prefactor from solving the quadratic as ${\bar g}_{1/2}^2 =(1/4)( 1 +\sqrt{1-12 \Lambda^{\prime}})$.

There is  also the possibility of the damping and Landau terms balancing, to give in a narrow $\tau$ region,  a small final tail ${\bar g} \sim 1/t << 1$ with  $\alpha =1$,  but after inaccessibly long crossover times \cite{R25}.

\subsubsection{$\tau < \tau_g < 0$  quench range} 

For deep quenches well below the $\tau =0$  spinodal, simulations show a peculiar curvature flattening $g(t) \rightarrow g_0$, or the exponent  $\alpha =0$ of Fig 10, corresponding to  the possible DW glass \cite{R26} of Fig. 4(c) . A similar ``coarsening arrest'' has been considered elsewhere \cite{R22}. We here suggest a toy model, to provide some understanding.
 
\subsection{Coarsening arrest: a toy model}
 
 The closure approximation (5.14c) had kept only the leading term in $g \xi_0$, approximating ${G^{(\mu_{L})}} (\bar R, g \xi_0)\simeq G^{(\mu_{L})}(\bar R,0) = -{\bar f}_0 ~G (\bar R)$,  as is reasonable for an asymptotically vanishing curvature.  However, if  for deep quenches  the curvature is constant, then with
 higher terms,  $G^{(\mu_{L})} (\bar R, g \xi_0)  \simeq \left[-{\bar f}_0   +{\bar f}_1~  (g\xi_0) - {\bar f}_2 ~(g \xi_0)^2 \right]~G$, where ${\bar f}_n$ are constants. For (6.2) in  the damping-dominated regime,

$$
 K_{10} \left\{\frac{-\dot g}{g}\right\} g^2  = \left[J_{30} - X_{40} g + X_{50} g^2 \right] g^3   + J_{40}(\tau)  g^4 $$
$$= J_{30} g^3 + \left\{J_{40} (\tau) - X_{40 }\right\} g^4 + X_{50} g^5 ~~(6.10)
$$
where $X_{40} \sim {\bar f}_1, X_{50} \sim {\bar f}_2$ are the {\it ex}tra coefficients, assumed for simplicity to be positive, $\tau$-independent constants.

\begin{figure}[ht]
\begin{center}
\includegraphics[height=3.5cm, width=8cm]{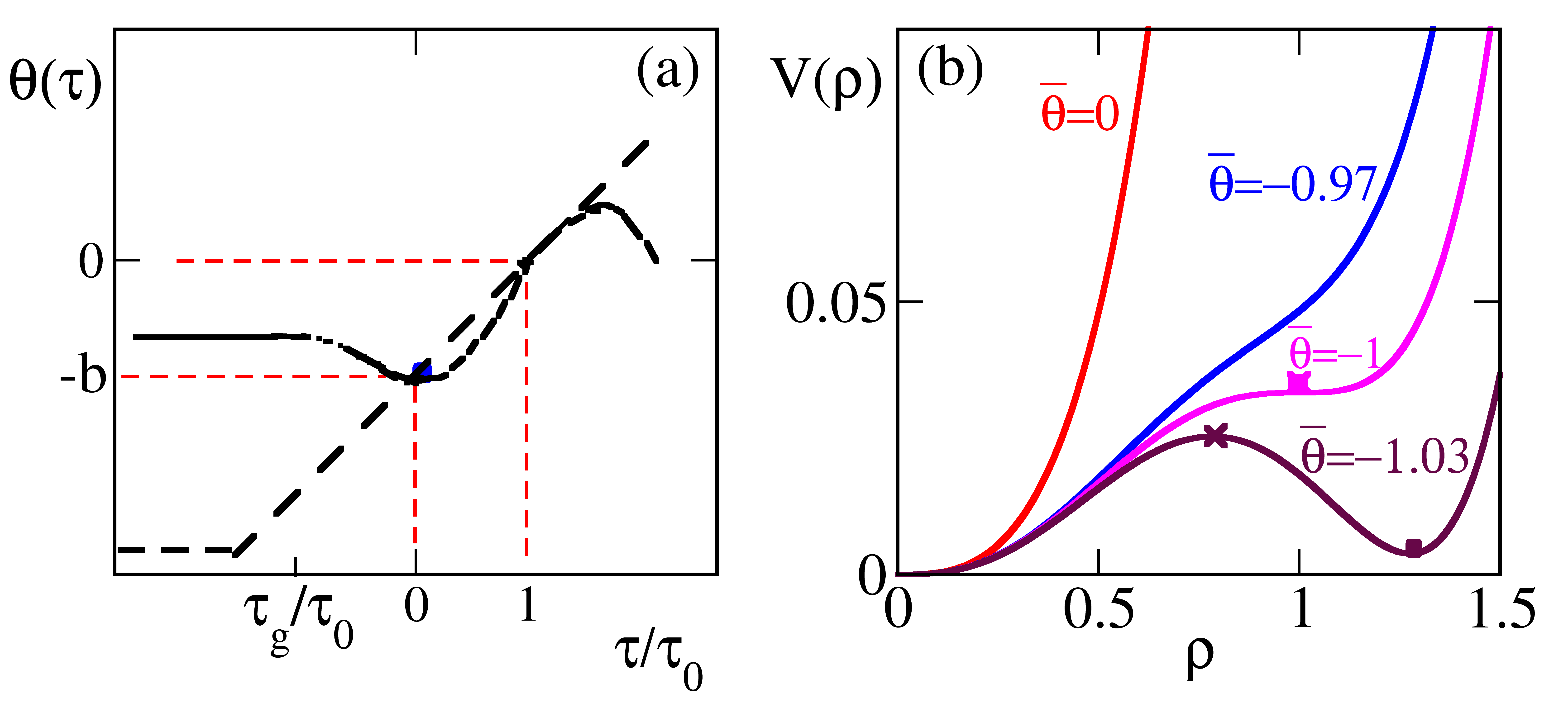}
\caption{(Color online) {\it Curvature Potential:}  (a) Schematic plot of parameter $\theta (\tau) = 1-J(\tau_0) / J(\tau)$ versus $\tau$, showing linear approximation and special temperatures $\tau = \tau_0, 0, \tau_g$. (b) Plot  of curvature potential $V(\rho)$ versus scaled curvature $\rho$, 
for various  decreasing values of the scaled variable ${\bar \theta } (\tau)$. 
Note onset of a metastable minimum in curvature, below $\bar \theta = -1$,  at a coarsening arrest onset temperature $\tau_g$. }
\label{Fig.13}
\end{center}
\end{figure}

Figure 15(c) of Appendix B shows $J_{40} (\tau)$ has a dip near  $\tau =0$.  Just above this, the falling $J_{40} (\tau)$ could cross the constant at some positive  $\tau =\tau_0 > 0$ where  $J_{40}(\tau_0) = X_{40}$.  Then doing scaling as before to absorb $|J_{30}|, |J_{40} (\tau)|$,  and with the scaled version of the coefficient $X_{50}$  factor written without subscripts, as  ${\bar X}$, 

$$  {\bar g}^2 (- {\dot {\bar g}}/ {\bar g}) = {\bar g}^3 +\theta (\tau) {\bar  g}^4 + {\bar X} {\bar g}^5. ~~(6.11)$$
where 
$$\theta(\tau) \equiv 1 - J_{40} (\tau_0) / J_{40} (\tau) \simeq b(\tau / \tau_0 -1) ~ (6.12)$$
and $\theta(\tau_0) =0$.
 
 Drawing on the $J_{40} (\tau)$ behaviour of Fig. 15(c), we assume a $\theta$ versus $\tau$ curve as in the schematic of Fig. 13(a), 
 and for ease of discussion, assume linearity around $\tau_0$, as in (6.12),  followed by a low-temperature levelling (dashed curve). 
 The slope $b$,  in terms of the $J_{40}(\tau)$ values at $\tau = \tau_0, 0$, is 
$b = (J (\tau_0)/J(0)) -1 >0$. 

Forces in (6.11) vanish at  the usual zero-curvature ${\bar g} =0$ final value. However, for $\tau/\tau_0 < 1$, i.e., for $\theta < 0$,  the net forces  can also vanish at a {\it nonzero}, metastable curvature ${\bar g}_0$.
Absorbing $\bar X$ by defining scaled curvatures and temperature deviations, 

$$\rho \equiv g \sqrt {\bar X} ; ~ {\bar \theta } (\tau)= \theta  (\tau) / 2 \sqrt {\bar X},  ~ (6.13)$$

we find (6.11) becomes
$$\dot \rho =   \frac{-\rho^2}{ \sqrt {\bar X}} \left[ 1+ 2 {\bar \theta} \rho + {\rho ^2} \right] \equiv \frac{-1}{\sqrt {\bar X}}\frac{ \partial V}{ \partial \rho} 
,~ (6.14)$$

where $V$ is an  effective curvature potential 
$$V (\rho, {\bar \theta})  = \frac{\rho^3}{3} + \frac{{\bar \theta} \rho^4}{2} +\frac{\rho^5}{5} ~~(6.15)$$
with maxima/ minima at roots
$$\rho_{\pm} (\tau)  = - {\bar \theta (\tau) } \pm \sqrt{ {\bar \theta (\tau)}^2 -1} ~~(6.16)$$
provided ${\bar \theta} < 0$ and ${\bar \theta}^2 > 1$.
The roots are  real only for (sub-spinodal) deep quenches $ \tau <\tau_g$, below  the glassy or `coarsening-arrest' temperature $\tau_g <0$, 
 defined by ${\bar \theta} (\tau_g) \equiv -1$. Here,  

$$\tau_g  /\tau_0 = - \left[(\sqrt{4 {\bar X}}/b) -1 \right] < 0, ~~(6.17)$$ 

and  $\rho (\tau_g) =1$. See Fig. 13(a). 

The curvature potential $V$  is plotted in Fig. 13(b), showing manifestly metastable minima.  
To check that  parameters are reasonable and obey required constraints, we draw on Fig. 15(c) to estimate values as  
$\tau_0 \simeq +0.025, J(\tau_0) \simeq 10, J(0) \simeq 1$ so that $b \simeq 10 >0$.  
Taking ${\bar X}= 100$, one has  $\tau_g \simeq -0.025 < 0$; and $\theta (\tau_g) \simeq -20$. The trapped curvature  is then $g_0 (\tau_g) \simeq 0.1$, comparable to  the flat value of Fig. 10.   

  The intermediate-time curvature decay  towards the metastable value $g_0$,  is  dominated by  the highest power,  $\dot {\bar g}  \simeq - {\bar X} {\bar g}^4$, that yields ${\bar g} \sim 1 /{\bar  t}^{1/3}$, or $\alpha = 1/3$, preceding the curvature flattening, as is indeed the case in Fig. 10.
   The ($d$-independent)  toy model thus  explains the relevant coarsening-arrest features seen in Fig. 10 for 2D, and in Fig. 12 for 3D.
 
\section{Discussion and Future Work}

Dynamical scaling is found in numerical simulations for  martensitic models with first-order transitions and Bales-Gooding dynamics. The coarsening exponent values include $\alpha =2/3, 1/2$ for intermediate and long times. For deep quenches there is some indication $\alpha= 1/3$ can occur, before an $\alpha =0$  value of coarsening arrest. 

The simulation exponents are theoretically understood through a curvature kinetics, can be generally derived as follows. 
i) Derive the dynamics of the two-point OP-OP correlation function, from a given OP-dynamics.
ii) Insert a dynamical scaling form, as an ansatz solution, with partial time derivatives now yielding total time derivatives of the curvature,  multiplying  space derivatives of the scaled correlation function. 
iii) Make approximations that (a) treat the chemical potential-OP correlation as a DW-OP correlation and (b) spatially average the internal  DW profile to yield a two-point OP-OP correlation, providing closure of the hierarchy.
iv) Evaluate coefficients at a dominant curvature front, yielding a  characteristic curvature kinetics for the  given OP-dynamics.
v)  Balance kinetic and force terms, to find powerlaw contributions, and their exponents.

We will elsewhere study  the effects of compatibility-induced power-law anisotropic interactions. Since the Fourier kernels are scale-independent,  dynamic scaling could  again  hold. Further work could study multicomponent  martensitic order-parameter dynamics with $N_{\rm OP}=2,3$ and $N_V=3,4,6$; and for both 2D and 3D.  

More generally, the curvature kinetics method could be tried out on other models such as  binary fluids, where different sequential exponents $\alpha = 1/3,1,2/3$ also  occur \cite{R1}. Of course, in the case of fluids, we have to deal with two coupled equations for the composition and velocity fields.\\

{\bf Acknowledgements}:  It is a pleasure to thank Prasad Perlekar and Surajit Sengupta  for useful conversations. NS acknowledges support from the University Grants Commission, India for a Dr. D. S. Kothari Postdoctoral Fellowship.

\vspace*{2cm}

\noindent{\bf Appendix A: Closure approximation for Correlation Dynamics}\\ \\ 

The  DW-OP correlation of (5.14a) is
$$  C^{(\mu_{L})} = -<  f_0 (e) ~e(\vec r,t) ~e(\vec {r^{\prime}},t)>~, ~~~~(A1)$$
where the  factor $f_0 (e)$ vanishes for the equilibrium, uniform  OP, $f_0 (\bar \varepsilon) =0$, and  is  nonzero only  in a region of order $\xi_0$ around the domain walls, i.e., in a relative volume $(\xi_0/L)$.  In a closure approximation, we spatially average at each domain wall, the factor as $<f_0 (e(x))> \equiv {\bar f}_0 (T)$ so
$$
{C}^{(\mu_L)} (R,t)  \simeq - (\xi_0/L)  {\bar f}_0 (T) G(R/ L) .~~~(A2)
$$

Note that \cite{R1} the total chemical potential around a spherical domain wall is  a constant $\mu <0$ inside, and $\mu \sim -\sigma/R= - \sigma / {\bar R} L$  outside, where $\sigma$ is the surface tension. Hence for fixed ${\bar R}$, one  also has effectively, $\mu \sim 1/L$.

Now we estimate the average ${\bar f}_0$ for a domain wall.
For  the CH case and a double-well scaled  Landau potential, $f_L =  -e^2 + e^4 /2$,
so $f_0 (e) =1-e^2, ~~$ as plotted versus $e$ in Fig. 14(a). In a direction perpendicular to the domain wall and a thickness $\sim \xi_0$, we take  a linear profile  ${ e} = x$ for $| x| < 1$;  
and flat at ${ e} (x)= \pm 1$  for $| x | > 1$.  Then spatially averaging,   $<\bar x^2> =1/3$  and
$${\bar f}_0 =(2/3) ~~~~~~~~~~~~~~~~~~~~~~~~~~~~~~~~~~~~~(A3) $$
 is constant,  as in Fig. 14(b). Thus, for the CH case,  $J_3 \sim {\bar f}_0 >0$ is always positive, or $\sigma_3 =+1$ as used in the text.

\begin{figure}[ht]
\begin{center}
\includegraphics[height=6.8cm, width=8.5cm]{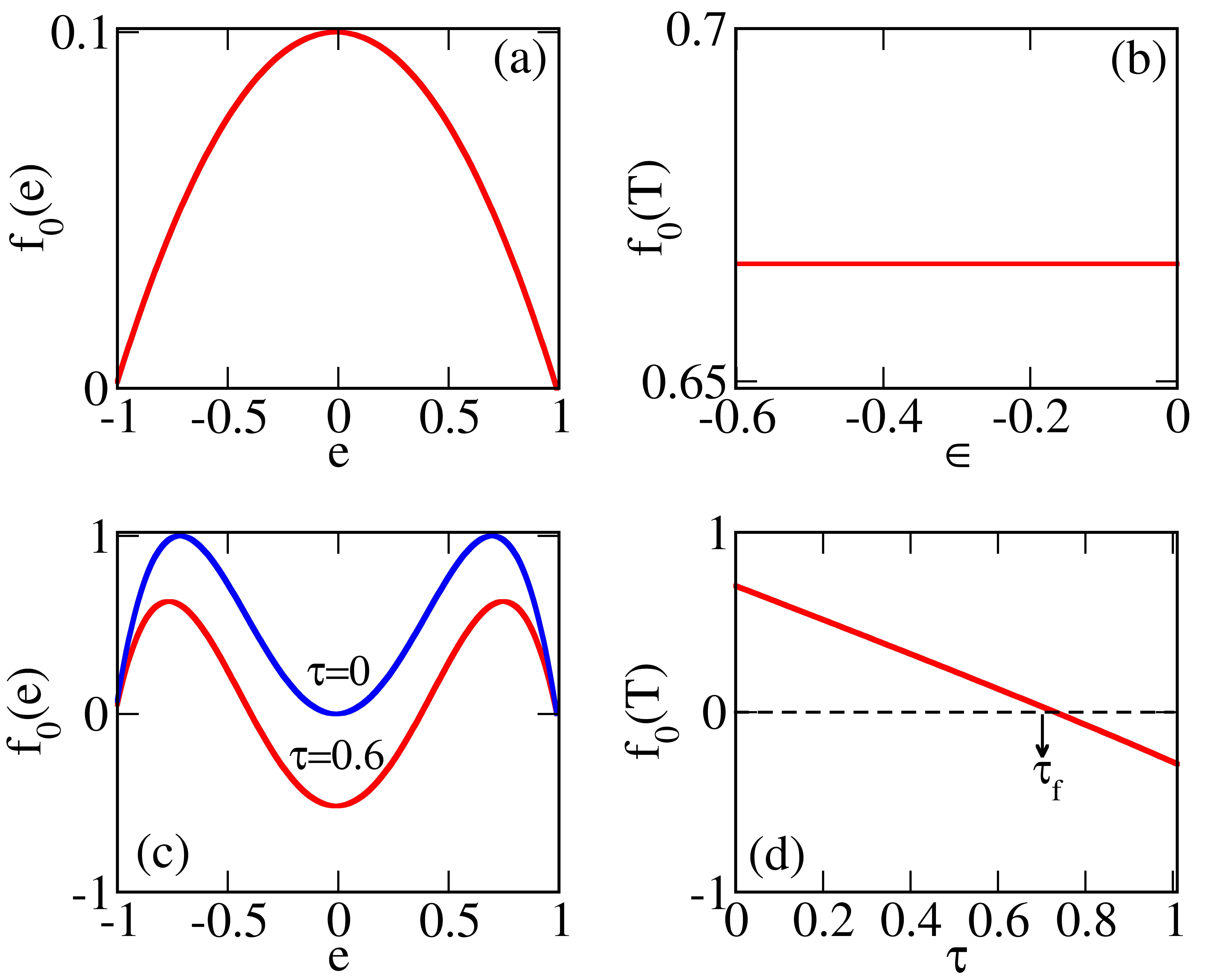}
\caption{(Color online) {\it Evaluation of prefactor ${\bar f}_0$}: First row: (a) Plot of CH case  $f_0 ( e)$ versus $ e$, showing it has a maximum at $e=0$, at the centre of a domain wall, and is zero in the bulk on either side. (b) CH case average ${\bar f}_0 (T)$ versus $\epsilon$ showing constant positive value. Second row: (c) Plot of  BG case $f_0 ( e)$ versus $e$ for $\tau =0, 0.6$, showing a double peak, with a minimum at $ e=0$, that can give a negative $f_0 (e)$;  (d) BG case  average 
${\bar f}_0 (T)$ versus $\tau$ showing change of sign  from negative to positive, on cooling through  $\tau ={\tau}_f$. 
 The approximation gives $\tau_f = 0.7$, while the simulation suggests from exponent-change, that $\tau_f \simeq  0.3$.  }
\label{Fig.14}
\end{center}
\end{figure}
 
For BG dynamics, and a triple-well scaled Landau potential,   $f_0 (e) = 3[1- e^2)(e^2 - \eta_{\rm sc} (\tau))]$ as  plotted versus $e$  in Fig. 14(c), where $\eta_{\rm sc} \equiv \tau / 3 {\bar \varepsilon}^4$. We take linear profiles $ e = x$ as before, for simplicity (although the martensitic profiles actually are different \cite{R14}), to obtain

$${\bar f}_0 = 3 [ (1+ \eta_{\rm sc}) < x^2> - <  x ^4> - \eta_{\rm sc}], ~(A4)$$
or
$$ {\bar f}_0 (\tau) =(2/5)  [ 1- 5 \eta_{\rm sc} (\tau) ].~~(A5)$$ 
From Fig. 14(d),   ${\bar f}_0 (\tau)$ changes  sign  from negative to positive on cooling through  some $\tau = \tau_f$. The temperature dependence of $f_0 (T)$ comes from the first-order nature of the Landau free energy $f_L (e)$.   A linear form is
$$ {\bar f}_0 (\tau) =(2/5) [ 1- \tau/ \tau_f  ].~~(A6)$$ 
The exponent $\alpha =2/3$ is supported for $\tau >\tau_f$ when ${\bar f}_0 < 0$, 
and simulations find this exponent for $\tau < \tau_f \simeq +0.3$. However, the above  mean-field-like approximation yields a higher value,  $\tau_f \simeq + 0.7$. \\ \\ \\ \\ \\

{\bf Appendix B: Coefficients of curvature kinetics}\\ \\

The coefficients $J_3 (\bar R), J_4 (\bar R), K_1 (\bar R)$ are evaluated at some dominant scale ${\bar R}_0$. 
The scaled function $G(\bar R)$ is fitted to a hexic-exponential function
$$G(\bar R) = \left[1 + {\sum_{\ell = 1}}^6 b_{\ell} {\bar R}^{\ell} \right] e^{- \lambda {\bar R}} ~~(B1)$$
from the origin at $\bar R =0$ to $\bar R =10$. We choose ${\bar R}_0$ as the non-stationary inflection point where $G^{\prime \prime}({\bar R}_0) =0$, while 
$G^{\prime} ({\bar R}_0) \neq 0$. 

The CH  value is  ${\bar R}_0 = 4.4$, independent of temperature, as expected from the OP-scaled CH dynamics with a second-order transition. The coefficients are positive, $J_{30} =+ 0.03; J_{40} =+0.25$, so the signs are $\sigma_3 = \sigma_4 = +1$.

The BG value   ${\bar R}_0 (\tau)$ is temperature- dependent through  the residual $\eta_{\rm sc} (\tau)$ of (2.16) in the OP-scaled BG dynamics with a first-order transition. Figure 15(a) shows the values of ${\bar R}_0$ versus $\tau$, and Fig. 15(c) shows the coefficients evaluated at  this ${\bar R}_0 (\tau)$. Since $K_{10} > 0, J_{40} > 0$, the signs are  $\sigma_1 = \sigma_4 = +1$ always. From Fig 15b, the sign of $J_{30} \sim f_0 (T)/ {\bar R}_0 (\tau)$  is negative for $\tau > \tau_f$ (supporting an exponent $\alpha =2/3$), but changes sign to positive on cooling through  $\tau_f$. These results are used in the curvature kinetics of the text. 

The text also has a toy model for coarsening arrest \cite{R22} with the $g^4$ effective-coefficient dependent on $J_{40} (\tau) -\bar X$ where $\bar X$ is a constant. Note the Fig 15a local maxima in ${\bar R}_0$ at $\tau =0$ and $0.3$, show up in the Fig 15c curves of  $ J_{40} (\tau) \sim 1/{ \bar R}_0 (\tau)$, as local minima.
 Thus the fall of $J_{40} (\tau)$  for temperatures just below the $\tau =0$ spinodal, can make the  effective $g^4$ coefficient  $J_{40}(\tau) -\bar X $ go negative at low temperatures, supporting a metastable glassy state of trapped curvature.

 \begin{figure}[ht]
\begin{center}
\includegraphics[height=6.5cm, width=8cm]{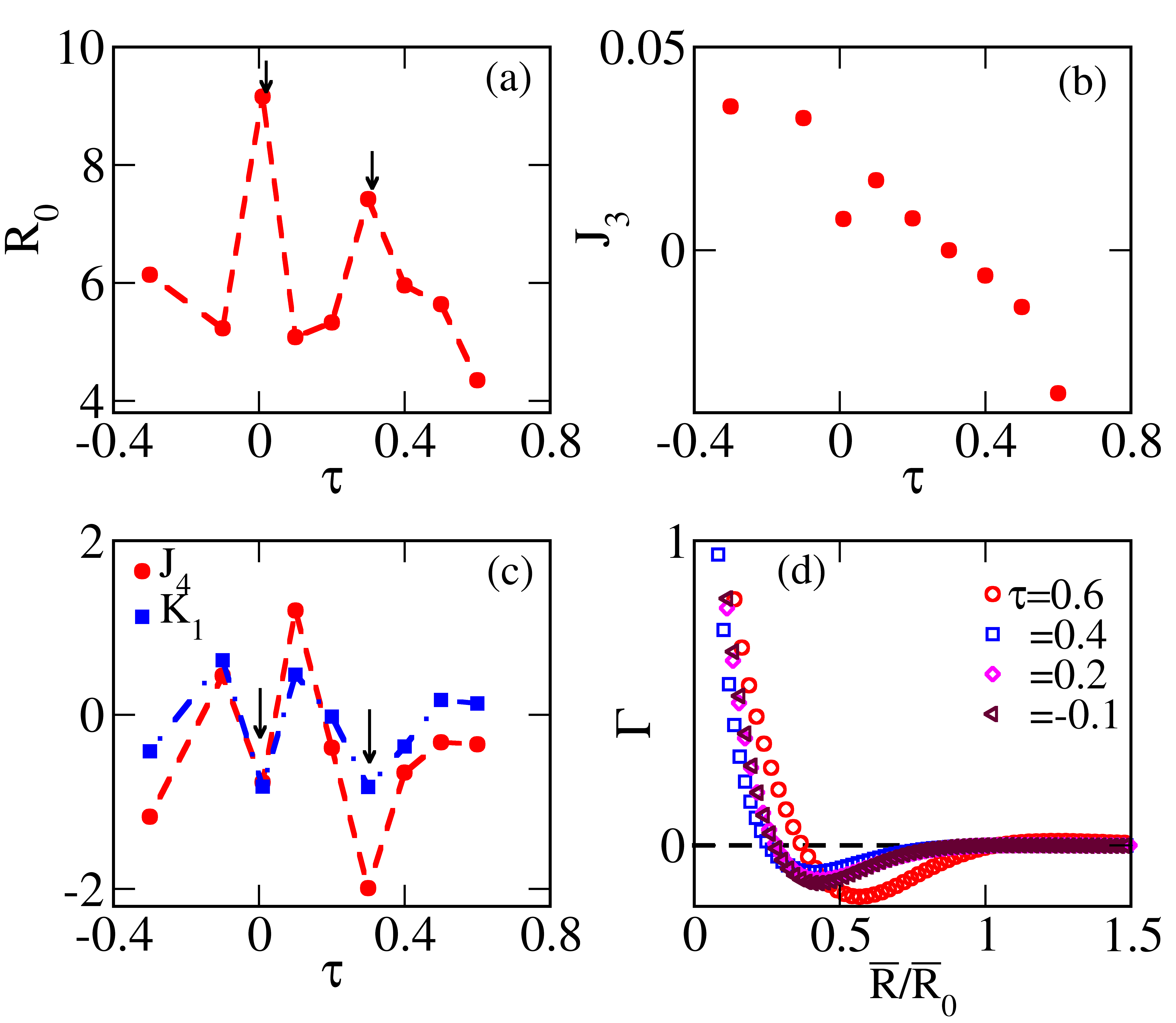}
\caption{(Color online) {\it Signs of  BG case curvature-dynamics coefficients: }  a) Plot of coefficient-evaluation choice ${\bar R} _0 (\tau)$ versus $\tau$ from data, with dashed lines as guides to the eye. Arrows mark local maxima near $\tau =0$ and $\tau= \tau_f =0.3$.  b) Plot of $J_{30} \sim {\bar f}_0 (T)/ {\bar R}_0 (\tau)$ versus $\tau$, showing the change in sign at $\tau_f \simeq 0.3$; 
c)Log-linear plot  of  (positive) coefficients $\log_{10}(J_{40})$,  $\log_{10}(K_{10})$ versus $\tau$. Arrows mark local minima near $\tau =0, 0.3$. d) Plot of domain-wall correlation $\Gamma (\bar R)$ versus $\bar R/ {\bar R}_0 (\tau)$, showing that it flattens to zero,  close to our  inflection-point choice of ${\bar R}/ {\bar R}_0 (\tau)$.  }
\label{Fig.15}
\end{center} 
\end{figure}

An alternate choice of $\bar R$ for coefficient evaluation, is where the domain-wall correlations fall to zero. As domain walls carry  nonzero  OP gradients, we define the gradient-gradient OP correlations scaled in the curvature as 
$$\Gamma (\bar R) \equiv g^{-2} \langle \nabla_{\vec r}~ e(\vec r,t). \nabla_{\vec {r^{\prime}}}~ e(\vec {r^{\prime}},t) \rangle = g^{-2} \nabla_{\vec r}.\nabla_{\vec {r^{\prime}}} G(\bar R)$$
$$ =-{\hat D} G(\bar R)= -[G^{\prime \prime} + (d-1) G^{\prime}/ {\bar R} ].~(B2)$$
This is a measure of DW correlations during coarsening, and also appears in the correlation dynamics of (5.9a), (5.9b). 
Fig 15d shows that $\Gamma (\bar R)$ flattens to zero close to ${\bar R}/{{\bar R}_0}=1$, so this alternative choice gives the same coefficient signs, as our $G$-inflection choice. \\ \\

{\bf Appendix C: Coarsening exponents} \\ \\

We here outline the procedure for numerically extracting, from  curvature falloffs, the exponent values given in the text.

A possible diagnostic for whether $g(t)$ has a powerlaw decay component $\sim 1/t^\alpha$ is to plot $t^\alpha g(t)$ versus $t$. It will flatten, where $\alpha$ is the most prominent contribution, and fall (or rise) as $t^{\alpha -\beta}$, where another exponent $\beta$ contributes more substantially. Fig 16a shows the variable

$$ Y_\alpha = (t {\bar \varepsilon}^4)^\alpha ( g / {\bar \varepsilon}^2), ~~~(C1)$$
plotted versus $(t {\bar \varepsilon}^4)$ for the test or trial values $\alpha= 2/3, 1/2$. This shows clear signatures of  single powerlaw decay contributions, with actual exponents close to these trial values. 
A supporting  width-diagnostic for the time windows, is $d \log g / d\log t$ versus $t$ (not shown): although the data is noisy, it also shows flat regions as in Fig 16a. 
Fig 16b shows linear fits in log-log plots within these single-powerlaw, dominance windows, that yield the actual, numerically fitted exponents.

\begin{figure}[ht]
\begin{center}
\includegraphics[height=3.5cm, width=8.5cm]{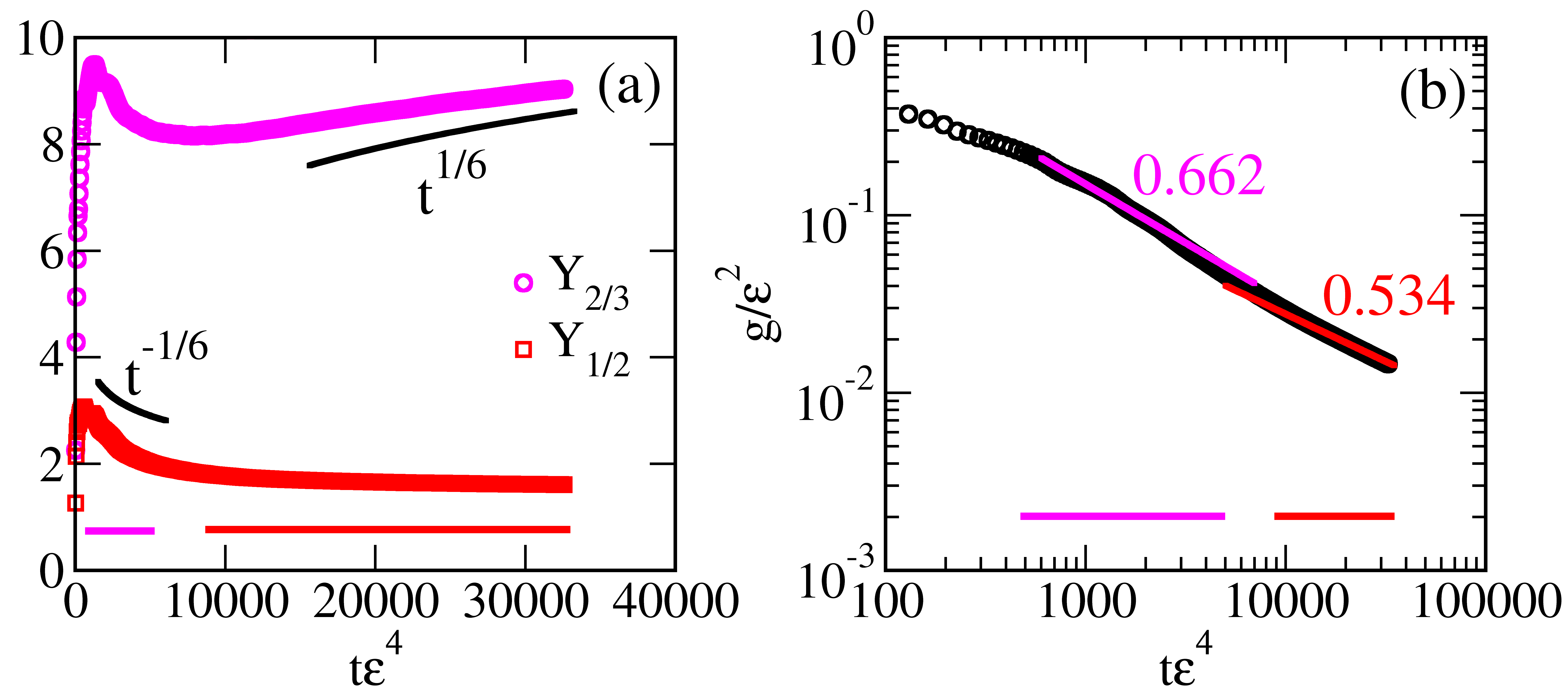}
\caption{(Color online) {\it Intermediate and late time BG exponents for $\tau = 0.2$ :} (a) Plot of data for  $Y_\alpha$ defined in the text vs $t {\bar \varepsilon}^4$, for test values $\alpha = 2/3, 1/2$. Flat regions are from dominance of  single power-laws, with exponents close to these test values. Horizontal bars denote the time windows  taken, for numerical fits. (b) Plot of $\log_{10} g/{\bar \varepsilon}^2$ vs $\log_{10} t {\bar \varepsilon}^4$ showing lines numerically fitted, within the time windows (see text).}
\label{Fig.16}
\end{center}
\end{figure}

With this procedure, in the dominance time-windows, the  exponent mean values and standard deviations,  for $\tau =0.6,~0.5,~0.4,~ 0.3,~ 0.2, ~0.1, ~0.05$  are found  to be respectively 
$ \alpha = 0.619 \pm 0.012; ~~ 0.680  \pm 0.018;~~ 0.688 \pm 0.014; ~~0.683 \pm; ~~0.662 \pm 0.015;  ~~0.636 \pm 0.013; ~~0.649 \pm  0.026$, which are all close to $2/3$. For $\tau = 0.3,~ 0.2,~ 0.1,~ 0.05$ the exponents are found to be  $\alpha = 0.555 \pm 0.03;~~ 0.534 \pm 0.02; ~~0.543 \pm 0.01; ~~0.497 \pm 0.01$,  which are all close to $1/2$.  The exponents are expected to be $\tau$-independent, and a simple arithmetic average yields  $\alpha = 0.661 \pm 0.017; ~~\alpha = 0.531 \pm 0.016$. Keeping two significant figures for consistency, 
$$\alpha = 0.66 \pm 0.02, ~~~{\text and}~~~ 0.53 \pm 0.02.~~~~(C2)$$
 For deep quenches of Fig 10, the other exponents seen are $\alpha = 0.35 \pm 0.03; ~~\alpha = 0.001 \pm 0.002$, which are close to $1/3$ and $0$. See Sec. VI on coarsening arrest.


\begin{thebibliography}{2004}
\bibitem{R23}
The evolutions from random dilute seeds in a zero-OP background depend on both the temperature and the seed size, as is reasonable for a first-order transition with a critical droplet radius for expansion. A surrogate-droplet parametrization  of complex-texture energies was introduced in a discretized-strain Monte Carlo context \cite{R18}, with the Hamiltonian energy $H(t)$  taken as proportional to  simple energy expression of a surface plus bulk energy in  2D as  $ f_L (T)R(t)^2 + R(t)$. For $t=0$ (only), the energy parameter  $R(t)$ has the geometric meaning of an initial seed size $R(0)$. There is a critical parameter $R_c(T)$ that for $A_1=0$ is defined as $R_c \equiv  2 \xi_0^2 {\bar \varepsilon}^2/ |f_L (\tau)|$, that diverges at $\tau = 1$.  The surrogate-droplet energy  versus $R(0)/R_c (T)$, is a single inverted parabola for all temperatures. Our continuous-strain dynamics simulations also fall on such a curve, as a bench-mark.
\bibitem{R24}
We multiply the dynamics  for $e({\vec r},t)$ by $e(\vec {r^{\prime}},t^{\prime})$ and set $\vec r=\vec R+\vec r_0$, $\vec {r^{\prime}}=\vec r_0$, to get an equation for $C(\vec R,t,t^{\prime})$ where $\vec R=\vec r-\vec {r^{\prime}}$. We then  take a dynamics  for $e(\vec {r^{\prime}},t^{\prime})$ and multiply by $e({\vec r},t)$, set $\vec r=\vec r_0$, $\vec {r^{\prime}}=-\vec R+\vec r_0$, to get an equation for $C(-\vec R,t,t^{\prime})$. Adding, and letting $t^{\prime} \rightarrow t$, we get a dynamics for the symmetrized OP-OP correlation $C_{sym}(\vec R,t)=\frac{1}{2}[C(\vec R,t)+C(-\vec R,t)]$ in terms of the symmetrized $\mu$-OP correlation ${C_{sym}}^{(\mu)}=\frac{1}{2}[C^{(\mu)}(\vec R,t)+C^{(\mu)}(-\vec R,t)]$, whose Fourier transform thus depends only on the real part $ {C_{sym}}^{(\mu)}(\vec k) \sim Re[\mu (\vec k,t){e(\vec k,t)}^{\ast}]$, and we drop the subscript.  On angular averaging, the correlation depends on $|\vec k|$ only.
\bibitem{R25}
From the unscaled curvature kinetics of (6.2), in the damping regime, a powerlaw falloff $g \simeq (K_{10}/ J_{40})^{1/2} / t^{1/2}$ could crossover to  an even smaller tail $g \simeq (K_{10}/ J_{30})/t  <<1$, at  very long times $\sim 10^5$,  much larger than the present holding time $t_h = 20,000$.
\bibitem{R26}
In Monte Carlo simulations \cite{R18}, (with compatibilty interactions),  there are transient zeros of the OP acting as dynamical catalysts, that facilitate domain wall rearrangements.
For $\tau$ well below the Landau spinodal, nucleation of these catalysts became less probable, and domain walls can form a glassy state, with trapped internal stresses.
In the present OP dynamics simulation, a similar  glass-like trapping seems to occur, for sub-spinodal quench temperatures. \\ \\
\end{thebibliography}
\end{document}